\documentclass[12pt]{article}
\usepackage{amssymb}

\usepackage{times}

\title{The C-value enigma and timing of the Cambrian explosion}

\author
{Dirson Jian Li$^\ast$ \& Shengli Zhang\\
\\
\normalsize{Department of Applied Physics, Xi'an Jiaotong
University, Xi'an 710049, China}\\
\normalsize{$^\ast$ E-mail:  dirson@mail.xjtu.edu.cn.} }

\date{}

\usepackage{graphicx}

\begin{document}

\baselineskip24pt

\maketitle

{\bf \begin{center} ABSTRACT \end{center}

The Cambrian explosion is a grand challenge to science today and
involves multidisciplinary study. This event is generally believed
as a result of genetic innovations, environmental factors and
ecological interactions, even though there are many conflicts on
nature and timing of metazoan origins. The crux of the matter is
that an entire roadmap of the evolution is missing to discern the
biological complexity transition and to evaluate the critical role
of the Cambrian explosion in the overall evolutionary context. Here
we calculate the time of the Cambrian explosion by an innovative and
accurate ``C-value clock''; our result (560 million years ago) quite
fits the fossil records. We clarify that the intrinsic reason of
genome evolution determined the Cambrian explosion. A general
formula for evaluating genome size of different species has been
found, by which major questions of the C-value enigma can be solved
and the genome size evolution can be illustrated. The Cambrian
explosion is essentially a major transition of biological
complexity, which corresponds to a turning point in genome size
evolution. The observed maximum prokaryotic complexity is just a
relic of the Cambrian explosion and it is supervised by the maximum
information storage capability in the observed universe. Our results
open a new prospect of studying metazoan origins and molecular
evolution.}

\newpage
\section*{\small INTRODUCTION}

\noindent The broad outline of Cambrian diversification has been
known for more than a century, but only in the post-genomic era have
the data necessary to explain the nature of the Cambrian explosion.
This problem originated in the disciplines of paleontology and
stratigraphy, while the debate about it may be as old as the problem
itself \cite{1}\cite{2}\cite{3}. Some ascribed the Cambrian
explosion to intrinsic causes, while others believe that it may have
been triggered by environmental factors. Innovative ideas exploded
in the past decade with new fossil discoveries and progress in
biogeochemistry, molecular systematics and developmental genetics
\cite{4}\cite{5}\cite{6}\cite{7}. However, we still need insights
from other fields such as genome size evolution, self-organization,
complexity theory and the holographic principle
\cite{8}\cite{9}\cite{10}\cite{11} to fully resolve this
long-running problem.

There is a profound relationship between the Cambrian explosion and
the C-value enigma. Why did so many complex creatures appear in the
late Neoproterozoic and Cambrian, but not earlier or later?  We
believe that the nature and timing of the Cambrian explosion can be
determined by the evolution of genome size (see the schematic in
Supplementary Figure 1). We invented a "C-value clock" to calculate
the time of the Cambrian explosion based on genomic data. The basis
of the C-value clock depends on the notion that the evolutionary
relationship can be revealed by the correlation of protein length
distributions and the genome size evolution can be taken as a
chronometer.

The start of our theory is a formula for evaluating genome size
(namely C-value) of different species. According to this formula,
major component questions of the C-value enigma can be solved and
the genome size evolution can be illustrated. Consequently, the
genome size evolution can be taken as an accurate chronometer to
study the macroevolution. We found a unique turning point in genome
size evolution and calculated the time of the turning point, which
corresponds to the Cambrian explosion. We believe that the Cambrian
explosion was essentially a major transition of biological
complexity when the prokaryotic complexity reached its maximum
value. We suggest that the biological complexity is supervised by
the maximum information storage capability in the observed universe.\\\\

\section*{\small RESULTS AND DISCUSSIONS}

\subsection*{}{\bf Genome size evolution.} Genome sizes vary
extensively in or between taxa. We found that the genome size $S$
can be determined by two variables: the noncoding DNA content $\eta$
and the correlation polar angle $\theta$. Hence we obtained an
empirical formula of genome size for any contemporary species:
\begin{equation} S(\eta,\theta)=s_0\
\exp(\frac{\eta}{a}-\frac{\theta}{b}),\end{equation} where
$s_0=7.96\times10^6$ base pairs (bp), $a=0.165$ and $b=0.176$ were
obtained by least squares based on the data of $S$, $\eta$ and
$\theta$ for $54$ species (see Supplementary Table 1 and 2). We also
obtained another empirical formula of gene number $
N(\eta,\theta)=1.48\times10^4\
\exp(\frac{\eta}{0.463}-\frac{\theta}{0.157})$ and the relationship
between non-coding DNA and coding DNA for eukaryotes $\log N_{nc} =
2.81 \log N_c -12.5$. The predictions of the formulae agree with the
experimental observations very well (Fig. 1a, 1b). The empirical
formula of genome size is the start of our theory, which can be
verified by many agreements between its predictions and experimental
observations (especially the detailed agreements, Fig. 1, 3 and 4).

The formula of genome size for contemporary species can help us
write down the formula of genome size evolution from $t=T_0=3,800$
million years ago (Ma) (the beginning of life \cite{12}) to $t=0$
(today). We introduced a function $s(t)$  to describe the overall
trend of the genome size evolution according to the distribution of
species in the $\eta-\theta$ plane. This is the main assumption in
our theory. We can distinguish two phases in genome size evolution
(Fig. 2a). In phase $I$, all the species in the lower triangle of
the $\eta-\theta$ plane are simple prokaryotes and their non-coding
DNA contents are low. In phase $I\hspace{-0.8mm}I$, all the species
in the upper triangle of  the $\eta-\theta$  plane are eukaryotes,
and the non-coding DNA content increased to the maximum value
$\eta^*$ . It is reasonable, therefore, to take the critical event
that divides the two phases as the Cambrian explosion.

Thus, we can obtain the formula of genome size evolution:
$s_{I}(t)=s_1\exp(t/\tau_1)$ for phase $I$ and
$s_{I\hspace{-0.8mm}I}(t)=s_2\exp(t/\tau_{2})$ for phase
$I\hspace{-0.5mm}I$, where $s_1=1.98\times10^7$ bp, $\tau_1=644$
million years and $s_2=1.65\times10^9$ bp, $\tau_2=106$ million
years (Fig. 2b). The result qualitatively agrees with the
straightforward (but a little coarse) estimation of genome size
evolution in Ref. \cite{13} in that (${\it i}$) both genome size
evolution increase exponentially (namely linearly in Fig. 2b) and
(${\it ii}$) there is a unique turning point in genome size
evolution for our result or for the estimate (Fig. 2b). As expected,
the dividing value of genome size in our theory
$s_I(T_c)=s_{I\hspace{-0.8mm}I}(T_c)=s_0$  agrees with the maximum
prokaryotic genome size in observation \cite{8}.

\subsection*{}{\bf Explanation of the C-value enigma.} The C-value
enigma is apparently concerned with the lack of correlation between
genome size and morphological complexity but profoundly with the
nature of the Cambrian explosion. According to the genome size
formula, we obtained some general properties of genome size
evolution, hence major questions of the C-value enigma can be
explained.

According to the genome size evolution formula, we can distinguish
two speeds of genome size evolution. In phase $I$, the genome size
doubled in about every $466$  million years on the whole. And in
phase $II$, the genome size doubled in about every $73$ million
years on the whole. So, the speed of genome size evolution for phase
$II$ (mainly non- coding DNA increasing) is much faster than that
for phase $I$ (mainly coding DNA increasing). The pattern of
exponential increment can be simply understood by the relation
$\Delta s(t) \propto s\ \Delta t$ for the two phases respectively.
The overall picture of the genome size evolution reflects the entire
roadmap of the biological complexity evolution, which is helpful to
understand the macroevolution.

The Cambrian explosion can help to account for the genome size
ranges in taxa. All phyla appeared almost simultaneously in the
Cambrian explosion. In the evolution, therefore, $\eta$ increases
from $\bar{\eta}$ to $\eta^*$ for each phylum (Fig. 2a). The genome
size in a phylum varies by about $\Delta = \lg \exp
\frac{\eta^*-\bar{\eta}}{a} \sim 2.4$ orders of magnitude (Fig. 3).
The history of a class is generally shorter than that of a phylum.
So the genome size range in a class is less than that in a phylum,
which varies by about $\delta = \lg \exp \frac{\Delta \theta}{b}
\sim 0.5$ orders of magnitude (Fig. 3), where the uncertainty
$\Delta \theta$ is estimated by $0.2$ (Fig. 2a). Furthermore, we can
explain the lack of correlation between genome size and
morphological complexity. The origin of phyla in the Cambrian
explosion related to the appearance of kernels of gene regulatory
networks, whose complexity varied notably. But the C-values of
species in different phyla did not vary notably \cite{5}. So the
discrepancy between genome size and eukaryotic complexity happened
from scratch (Fig. 3).

Three clusters of prokaryotes $C_{Gram-}$, $C_{Gram+}$ and
$C_{small}$ can be distinguished in the lower triangle of the
$\eta-\theta$  plane (Fig. 2a), where Gram negative bacteria, Gram
positive bacteria and bacteria with small genome size are in the
majority respectively \cite{32}. We evenly distributed $6038$ dots
(representing "species") in three symmetric areas enclosing
$C_{Gram-}$, $C_{Gram+}$ and $C_{small}$ in Fig. 4a (the same areas
with Fig. 2a). After projecting the three symmetric areas in  plane
by the non-linear transformation Eqn. 1, we obtained three
asymmetric areas $C'_{Gram-}$, $C'_{Gram+}$ and $C'_{small}$ in
$\eta-s$ plane in Fig. 4c. Finally, we obtained the prokaryotic
genome size distribution in Fig. 4b by counting the numbers of
species in each genome size section with identical width in Fig. 4c.

\subsection*{}{\bf Timing of the Cambrian explosion.}
The time $T_c$ for the Cambrian explosion can be calculated
according to the formula of genome size evolution. The function
$s_I(t)$ represents the coding DNA evolution. Its extrapolated value
$s_I(0)=s_1$ represents the size of coding DNA at present. And the
value $s_{I\hspace{-0.8mm}I}(0)=s_2$ represents the total genome
size at present. For the coding DNA content at present, we obtained
an equation between the experimental data and the theoretical
prediction
 $1-\eta^*=s_1/s_2$, where $s_1$ and
$s_2$ are functions of $T_c$. According to this equation, we have
\begin{equation} T_c=T_0
(1-(\frac{b}{1-\bar{\eta}} \ln
(1-\eta^*)+\frac{b}{a}\frac{\eta^*-\bar{\eta}}{1-\bar{\eta}}+
1)^{-1})\equiv f(\eta^*).\end{equation} This is the formula to
calculate the Cambrian explosion time by C-value clock, which
radically differs from molecular clock estimates (Fig. 2c) \cite{14}
\cite{15}. The value $\eta^*$ should be of the species whose $\eta$
is the largest and whose complexity is the greatest. The best choice
is no other than human: $\eta^*=0.988$ \cite{16} \cite{17}.
Therefore, we obtained the Cambrian explosion time
$T_c=f(0.988)=560$ Ma. Our result agrees with the fossil records
very well (Fig. 2d).

This main result of C-value clock shows that the Cambrian explosion
corresponds to a turning point in genome size evolution. It is for
the first time, to our knowledge, to successfully mediate timing of
the Cambrian explosion between paleontology and molecular biology.
Considering the sensitive relationship between $T_c$  and $\eta^*$ .
(Fig. 2d), it is remarkable to calculate almost the exact time of
the Cambrian explosion by the non- coding DNA content of human
genome. The subtle relationship  $T_c = f(\eta^*)$  indicates the
close relationship between the rapid expansion of noncoding DNA and
the cause of the Cambrian explosion. The genetic mechanism can give
us a clear and in-depth understanding of the Cambrian explosion.
Both development and evolution of the animal body plans should be
studied at the level of gene regulatory networks \cite{5} \cite{18}.
The appearance of genomic regulatory systems may be a prerequisite
for the animal evolution. And the phylum-specific or
subphylum-specific kernels of gene regulatory networks may explain
the conservation of major phyletic characters ever since the
Cambrian \cite{18}.

According to Eqn. 2, we obtained $\frac{\Delta
T_c}{T_c}=-75\frac{\Delta \eta^*}{\eta^*}+\frac{\Delta
T_0}{T_0}-5.2\frac{\Delta a}{a}+0.85\frac{\Delta
b}{b}-0.57\frac{\Delta \bar{\eta}}{\bar{\eta}}$. The error of $T_c$
in prediction, therefore, mainly comes from the parameter $\eta^*$.
If considering the uncertainty in human gene prediction, the error
of coding DNA content in human genome is about $10\%$ \cite{16}.
Hence we obtained that the value of $T_c$  in prediction ranges from
$502$ Ma to $560$ Ma. Even if the databases of complete genomes and
proteomes may expand much in the future, the parameters in Eqn. 1
would change slightly. So our main results in this paper will still
be valid. By the way, if choosing $T_c$ as the date of the earliest
known microfossils, i.e., $T_0=3,500$ Ma, the prediction would be
$T_c=516$  Ma. There is a notable discrepancy between the molecular
clock estimates and the fossil records \cite{15} \cite{33}
\cite{34}. Obviously, the C-value clock works better than the
molecular clocks for this problem. We can conclude that the C-value
clock estimate agrees with the fossil records in principle (Fig.
2d).

If comparing the time of evolution of life as a day, why did not the
complex life appear in the morning or in the afternoon but appear
around half past eight in the evening? In terms of the overall
picture of genome size evolution in Fig. 2b, we can explain why the
simple life had actually predominated on the planet for the first
$6/7$ time in the evolution. It is due to that the evolutionary
speed for non-coding DNA is much faster than that for coding DNA.The
Cambrian explosion can not happen in the first half of the period in
the evolution. The reason is that $s_1$  is always less than $s_2$
such that the turning point had to appear later than the time
$T_0/2$. Furthermore, it can be illustrated that the Cambrian
explosion must happen very late because $s_1$  is in fact much less
than $s_2$  at present, namely, the slope for the evolution of
non-coding DNA is much steeper than the slope for the evolution of
coding DNA (Fig. 2b).

\subsection*{}{\bf Nature of the Cambrian explosion.}
The formula of genome size evolution opens up an opportunity to
investigate the entire roadmap of evolution based on biological
complexity. It is observed that the biological complexity increases
faster and faster but not smoothly \cite{19} \cite{20} \cite{21}.
The pattern that mass extinctions followed by rapid evolutionary
radiations is widely considered to have fundamentally shaped the
history of life. But it is not the answer to the case of the
Cambrian explosion. The evolution is not only a mixture of
accidental events. The one with less perseverance can never spend
billions of years to assemble a jaguar by quarks! An overall
mechanism of the evolution is required to explain the Cambrian
explosion. The genome size evolution is just a problem on
macroevolution. In our theory, the function $s(t)$ represents not
only the trend of the genome size evolution but also the trend of
the biological complexity evolution because the prokaryotic
complexity is related to the genome size and the eukaryotic
complexity is related to the non-coding DNA content \cite{17}. The
turning point in genome size evolution implies that there was a
critical value of biological complexity in evolution, which is
supported by the fact that both the genome size and the complexity
of prokaryotes have never reached the size and complexity of
eukaryotes. The constraint of the prokaryotic complexity demands a
leap in biological complexity. As a result, the complex organisms
successfully bypassed this constraint during Cambrian.

Several attempts have been proposed to explain the maximum
prokaryotic complexity \cite{22} \cite{23} \cite{8}. Its existence
can be explained by the theory of accelerating networks \cite{24}.
It is suggested that prokaryotic complexity may have been limited
throughout evolution by regulatory overhead, and conversely that
complex eukaryotes must have bypassed this constraint by novel
strategies \cite{22} \cite{19}. We give another explanation based on
Kauffman's theory and the holographic principle \cite{9} \cite{11}
\cite{25}. The theory of self-organization provides deep insight
into the spontaneous emergence of order which graces the living
world \cite{9}. The prokaryotic complexity should be understood as a
dynamical system at the level of gene networks. So we can define
prokaryotic complexity by information stored in Boolean networks,
which is so immense that it can reach the maximum information
content $I_{univ}$ in the observed universe. Holographic bound in
physics imposes a strict limit on the biological complexity. The
information bridges between biology and physics \cite{26} \cite{27}.
We believe that the maximum prokaryotic complexity is constrained by
the upper limit of information storage capacity in our universe.
Hence the maximum complexity of accelerating networks in the above
explanation can be given concretely.

The Cambrian explosion of animal phyla radically differs from all
the other radiations such as the radiations of modern birds and
mammals in the early Tertiary, because it corresponds to the unique
critical event in the genome size evolution. The intrinsic reason of
genome evolution determined the Cambrian explosion, during which the
biological complexity leapt not only at the anatomical level but
also at the molecular level. The stability of the genomic system
became low before the Cambrian explosion because the old mechanism
of evolution was suffocated. At this critical moment, any extrinsic
factors were qualified to turn the evolution to a new direction.
Numerous complex animal body plans were destined to come at a
certain time. In contrast, the causes of other radiations were full
of uncertainty. The nature of the Cambrian explosion must be studied
in a broader context than before. The Cambrian explosion and the
origin of life were the most important events in the evolution from
nonliving systems to living systems. We believe that the C-value
enigma and the Cambrian explosion will help us uncover the intricate
mechanism in evolution. A multidisciplinary framework has been
established in our work to explain the Cambrian explosion (see
Supplementary Figure 1), which will shed light on the essence of
evolution.\\\\

\section*{\small METHODS}

\subsection*{}{\bf The definition of correlation polar angle $\theta$ and
its biological meaning.} The correlation polar angle indicates the
evolutionary relationship, whose role in the C-value clock is as
important as the role of sequence similarities in molecular clocks.
The correlation polar angle can be defined according to protein
length distributions, which helped in discovery of the formula of
genome size when we fortunately realized the relationship between
genome size $S$ and the correlation polar angle $\theta$. In the
followings, we define the correlation polar angle firstly. Then we
explain its biological meaning.

The protein length distribution is an intrinsic property of a
species, which is defined as a distribution (namely a vector)
${\mathbf D}=(D_1, D_2, ..., D_n,...)$: there are $D_n$ proteins
with length $n$ in the complete proteome of the species. Our data of
the protein length distributions are obtained from the data of $106$
complete proteomes in the database Predictions for Entire Proteomes
\cite{30}. The normalized vector of protein length distribution
${\mathbf d}$ is defined by the direction of vector ${\mathbf D}$:
$${\mathbf d} \equiv {\mathbf D} / \sqrt{\mathbf D \cdot \mathbf D} =
{\mathbf D} /\sqrt{\sum_{n} D^2_n}.$$ Because there are few proteins
longer than $3000$ amino acids in a complete proteome (Supplementary
Figure 2c), we can neglect them and set the length $3000$ as the
cutoff of protein length in the calculation. Hence both ${\mathbf
D}$ and ${\mathbf d}$ are 3000-dimensional vectors. Thus each
species corresponds to a point on the 3000-dimensional unit sphere
(Supplementary Figure 4a). The polar axis of the spherical
coordinates (Supplementary Figure 4a) can be defined by the
direction of the vector of the total protein length distribution of
the $106$ species (Supplementary Figure 2c) $${\mathbf Z}=\sum_{i
\in 106\ species} {\mathbf D}(i).$$ And we denote the normalized
vector of $\mathbf Z$ as the unit vector $\mathbf z$ of polar axis,
the corresponding point of which situates at the center of the swarm
of $106$ points on the unit sphere (Supplementary Figure 4a). The
correlation polar angle $\theta$ of a species is defined by the
polar angle of the corresponding vector of protein length
distribution: $$\theta \equiv \frac{2}{\pi} \arccos({\mathbf d}
\cdot {\mathbf z}),$$ where the factor $\frac{2}{\pi}$ is added in
order that the value of $\theta$ ranges from $0$ to $1$.

The biological meaning of the correlation polar angle can be
interpreted as the average evolutionary relationship between an
species and all the other species (Supplementary Figure 3b). The
less the value of $\theta$ is, the closer the average evolutionary
relationship is. This interpretation is based on the following two
considerations: (1) Let vectors ${\mathbf d}(i)$ and ${\mathbf
d}(j)$ correspond the protein length distributions of two species
$i$ and $j$ (Supplementary Figure 4a). The correlation between the
two protein length distributions can be defined by their inner
product $C_{ij}={\mathbf d}(i) \cdot {\mathbf d}(j)$. Hence we
obtain the correlation matrix $(C_{ij})$ (Supplementary Figure 3a).
We can see that the evolutionary relationship is closely related to
the correlation between the protein length distributions. The
correlation polar angle $\theta$ for species $i$ can be interpreted
as the average evolutionary relationship according to (compare
Supplementary Figure 3a and 3b):
\begin{eqnarray}\cos(\frac{\pi}{2} \theta) & = & \sum_{j
\in 106\ species} {\mathbf d}(i) \cdot {\mathbf D}(j)/ \sqrt{\mathbf
Z \cdot \mathbf Z}) \nonumber \\ & = & \sum_{j \in 106\ species}
\cos(\frac{\pi}{2} \theta_{ij})\ w(j) \nonumber,
\end{eqnarray}
where $\theta_{ij}=\frac{2}{\pi} \arccos(C_{ij})$ is the correlation
angle between two species and \\ $w(j)=\sqrt{{\mathbf D}(j) \cdot
{\mathbf D}(j)}/\sqrt{{\mathbf Z} \cdot {\mathbf Z}}$ is the weight
for species $j$ in the summation. (2) An auxiliary polar axis
${\mathbf z'}$ can be defined by another direction differed from the
polar axis. For example, we chose the direction corresponds to the
distribution in Supplementary Figure 2d, hence the auxiliary polar
angle is defined by $\phi \equiv \frac{2}{\pi} \arccos({\mathbf d}
\cdot {\mathbf z'})$ (Supplementary Figure 4a). Then the high
dimensional unit sphere (dim=3000) can be projected to a two
dimensional $\theta-\phi$ plane, where eukaryotes, archaebacteria
and eubacteria gather together in three areas respectively
(Supplementary Figure 4b) and the closely related species also form
clusters in the $\theta-\phi$ plane. So the correlation polar angle
is a useful tool to study the evolutionary relationship. The
conclusion is still valid if we choose other directions as the
auxiliary polar angle.

\subsection*{} {\bf Derivation of Eqn. 1: the genome size
$S(\eta, \theta)$.} We found that $\ln S$ decreases linearly with
$\theta$ (Supplementary Figure 5a) but increases linearly with
$\eta$ (Supplementary Figure 5b) on the whole. Hence, we wrote down
the relation:
$$ \ln S = \ln s_0 + \frac{\eta}{a}-\frac{\theta}{b}.$$ According to
the biological data of genome size, $\eta$ and $\theta$
(Supplementary Table 2), we obtained the empirical formula of genome
size Eqn. 1 and its coefficients $a$, $b$ and $s_0$ by least
squares. Similarly, we obtained the gene number formula
$$N(\eta,\theta)=n_0\ \exp(\frac{\eta}{a'}-\frac{\theta}{b'}).$$ The
value of $\eta$ varies little for prokaryotes in both formulae and
$b \approx b'$, so the genome size is approximately proportional to
the gene numbers: $$ \frac{S}{N} \approx \frac{s_0}{n_0}
\exp(\bar{\eta} (\frac{1}{a}-\frac{1}{a'})) = 842,$$ which is near
to the ratio in observation \cite{8}. But such linear relationship
is destroyed for eukaryotes because of the vast variation of $\eta$.

\subsection*{} {\bf The relationship between non-coding DNA
$N_{nc}$ and coding DNA $N_c$ for eukaryotes.} The average protein
length for eukaryotes is about 450 amino acids, so the logarithm of
coding DNA for eukaryotes is about $ \log N_c = \log (3 \times 450\
n_0) + \frac{\eta}{a'} - \frac{\bar{\eta}}{b'}$ according to the
gene number formula. And the logarithm of non-coding DNA is about $
\log N_{nc} = \log s_0 + \frac{\eta}{a} - \frac{\bar{\eta}}{b} +
\log \eta$ according to the genome size formula. So we have
\begin{eqnarray} \log
N_{nc} &=& \frac{a'}{a} \log N_c + \log s_0 - \frac{a'}{a} \log
(1350\ n_0) - \frac{\bar{\eta}}{b} +\frac{\bar{\eta}}{b'}
\frac{a'}{a} + \log \eta \nonumber \\  &\approx& 2.81 \log N_c -12.5
\nonumber,
\end{eqnarray}
where we let $\log \eta \approx \log 0.5$ in calculation. According
to the experimental observation (Figure 1 in Ref. \cite{31}), we
obtain the relationship $\log N_{nc} = 2.82 \log N_c - 12.8$ between
non-coding DNA and coding DNA for actual species on the whole if
choosing two points $(6.8, 6.4)$ and $(7.9, 9.5)$ in Figure 1 in
Ref. \cite{31} to determine the linear relationship. Our result
agrees with the experimental observation perfectly.

\subsection*{} {\bf The genome size evolution function $s(t)$.} We
can observe a right-angled distribution of the contemporary species
in the $\eta-\theta$ plane (Fig. 2a). The prokaryotes and the
eukaryotes are separated by the diagonal line $\eta = \theta$. An
underlying mechanism of genome size evolution is necessary to
account for the distribution. Some species originated earlier while
the other originated later. As a result, the distribution of species
in the $\eta-\theta$ plane has recorded the information of genome
size evolution. Hence we can write down the genome size evolution
function.

The prokaryotes situate around the horizontal line
$\eta=\bar{\eta}=0.115$, where $\bar{\eta}$ is the average of $\eta$
for $48$ prokaryotes (see Supplementary Table 1 and 2). According to
Eqn. 1, the trend of the genome size evolution for prokaryotes
increases when $\theta$ decreases. When $\theta$ is close to $1$,
there is few species because the genome size is too small as for the
contemporary species. On the other hand, the eukaryotes situate
around the vertical line $\theta=\bar{\eta}$ and the trend of their
genome size evolution increases when $\eta$ increases.

We introduced a function $s(t)$ to describe the overall trend of the
genome size evolution according to the right-angled distribution in
observation, whose turning point corresponds to the largest genome
size of prokaryotes (Fig 2a). It is reasonable to define that the
genome size evolution function $s(t)$ evolves leftwards along the
horizontal line $\eta=\bar{\eta}$ and consequently upwards along the
vertical line $\theta=\bar{\eta}$ in the $\eta - \theta$ plane. This
definition of genome size evolution function agrees not only with
the right-angled distribution of species in the $\eta - \theta$
plane but also with the trend of the genome size evolution from
small to large on the whole.

\subsection*{} {\bf Derivation of Eqn. 2: the Cambrian explosion
time $T_c$.} In phase I, $\eta(t)=\bar{\eta}$, and $\theta(t)$
decreases linearly from $1$ to $\bar{\eta}$, i.e., $\theta (t) = 1-
\frac{(1-\bar{\eta})(t-T_0)}{T_c-T_0}$. So we have $$s_I (t) \equiv
s_0\ \exp(\frac{\eta(t)}{a}-\frac{\theta(t)}{b}) = s_1 \exp
(t/\tau_1),$$ where $s_1 = s_0 \exp
(\frac{\bar{\eta}}{a}-\frac{T_c-\bar{\eta}T_0}{b(T_c-T_0)})$ and
$\tau_1 = \frac{b(T_c-T_0)}{1-\bar{\eta}}.$ Incidentally, we have
$s' \equiv s_I(T_0)=s_0 \exp (\frac{\bar{\eta}}{a}-\frac{1}{b})$.
And in phase I\hspace{-0.5mm}I, $\theta(t)=\bar{\eta}$ and $\eta(t)
= \eta^* - (\eta^*-\bar{\eta})t/T_c$. So we have
$$s_{I\hspace{-0.5mm}I} (t) \equiv s_0\
\exp(\frac{\eta(t)}{a}-\frac{\theta(t)}{b}) = s_2 \exp (t/\tau_2),$$
where $s_2 = s_0 \exp (\frac{\eta^*}{a}-\frac{\bar{\eta}}{b})$ and
$\tau_2 = \frac{a(-T_c)}{\eta^*-\bar{\eta}}.$ Finally, substituting
the expressions of $s_1$ and $s_2$ into the equation
$1-\eta^*=s_1/s_2$, we obtained Eqn. 2.

\subsection*{}{\bf Upper limit of the prokaryotic complexity.}
Boolean networks have for several decades received much attention in
understanding the underlying mechanism in evo-devo biology \cite{9}
\cite{35}. We define the network $N_L$  as a Boolean network whose
nodes are all possible protein sequences with the length less than
$L$. The size of state space of $N_L$ is $\sim 2^{20^L}$  in that
$N_L$ has about $20^L$ nodes. According to Shannon's theory, the
information stored in this network is $I_{net} \sim \log_2
2^{20^L}=20^L$ bits (Supplementary Figure 6). Types of prokaryotes
can be interpreted by attractors of the Boolean network $N_L$, which
are robust against perturbations in evolution \cite{35}. An actual
genome of an organism can be denoted by one point amongst the total
$\sim 2^{20^L}$  points in the state space of $N_L$. Based on the
consideration that the biological complexity should be evaluated at
the level of gene regulatory networks, the prokaryotic complexity
can be defined by the information $I_{net}$ stored in $N_L$. Its
value is much greater than the information stored in the genetic
sequences; the latter is not sufficient to measure the biological
complexity for overlooking the complexity at the level of gene
networks. This definition does not apply to eukaryotic complexity,
which may involve RNA regulations \cite{19}.

We can show that the constrained maximum complexity of unicellular
organisms can be explained by the upper limit of information stored
in the finite space. There was a great achievement in the knowledge
of fundamental laws in nature, which originated in the field of
quantum gravity \cite{36} \cite{37} \cite{25}. It claims that the
information storage capacity of a spatially finite system must be
limited by its boundary area measured in fourfold Planck area unless
the second law of thermodynamics is untrue. Consequently, we can
obtain the maximum information storage capacity in the observable
universe as $I_{univ} \approx 10^{122}$  bits \cite{38}, which is a
strict limit on the information content not only for physical
systems but also for living organisms. Let $I_{net}\sim I_{univ}$,
we obtained $L \sim 94$ amino acids, which dramatically corresponds
to the most probable protein length for prokaryotes (Supplementary
Figure 2b) \cite{39}. So the information stored in the prokaryotic
gene networks is so large as to be comparable to $I_{univ}$. Thus we
have demonstrated the equivalence between the prokaryotic complexity
and the information content $I_{univ}$  in our universe. We might
say that what kind of spacetime determines what kind of life. A
certain vast spacetime is necessary to accommodate the immense
information stored in life.

We are grateful to Hefeng Wang, Lei Zhang, and Yachao Liu for
valuable discussions. Supported by NSF of China Grant No. of
10374075.

\clearpage
\begin{figure}
\centering{
\includegraphics[width=89mm]{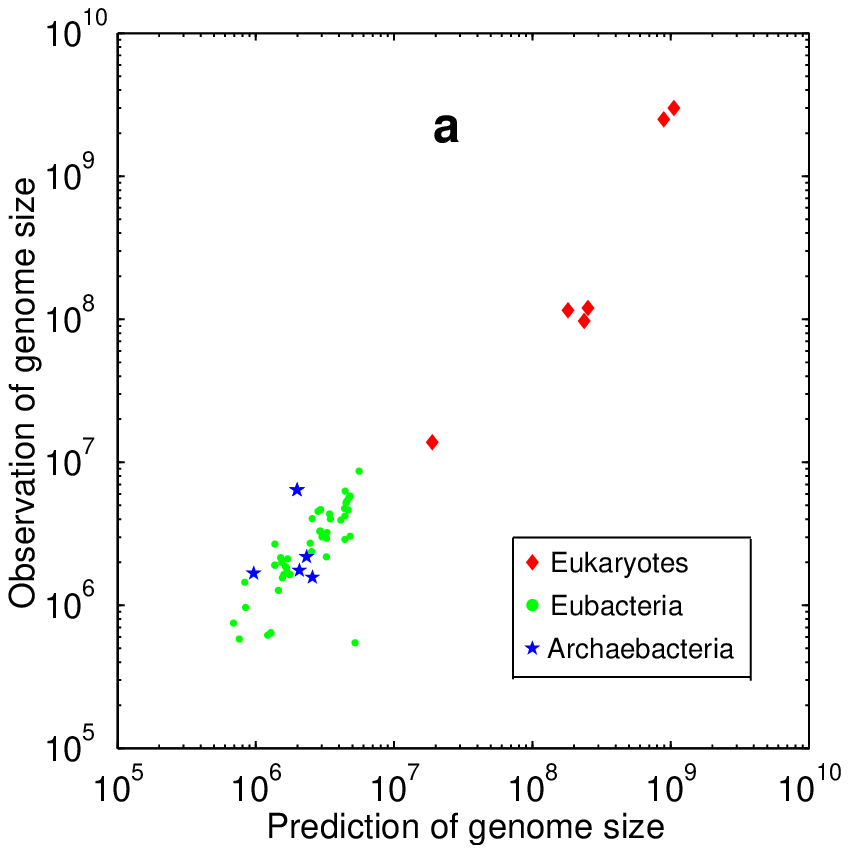}
\includegraphics[width=89mm]{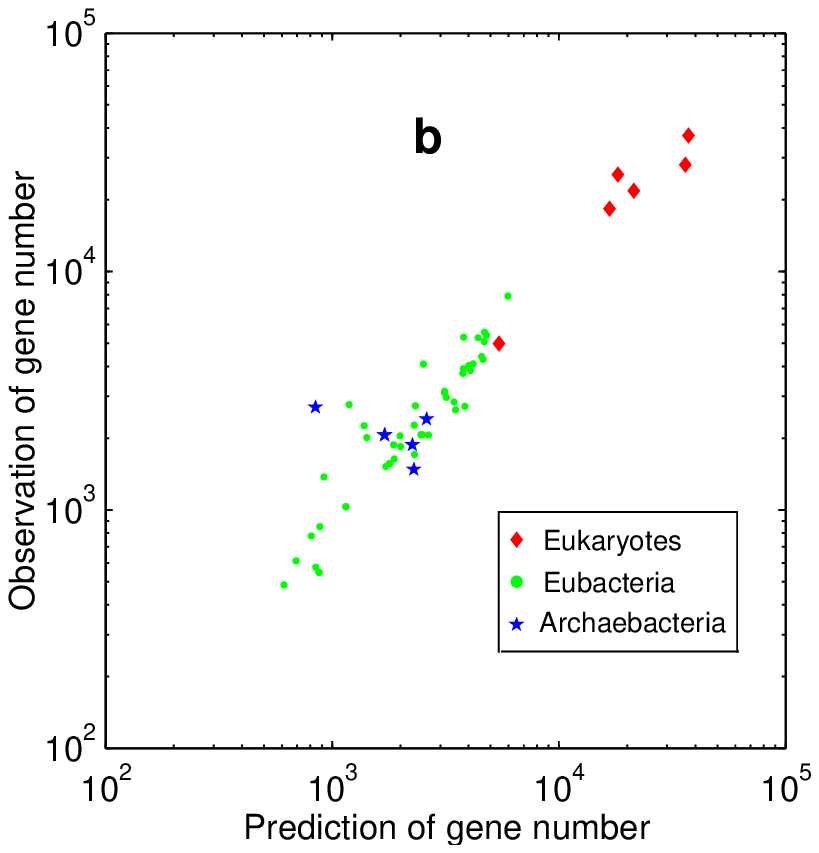}}
\label{fig} \caption{{\bf Comparison between predictions and
observations for genome size and gene number.} Our results quite fit
the experimental observations not only for prokaryotes but also for
eukaryotes. {\bf a,} Genome size (correlation coefficient
$r=0.974$). {\bf b,} Gene number (correlation coefficient
$r=0.976$).}
\end{figure}

\clearpage
\begin{figure}
\centering{
\includegraphics[width=75mm]{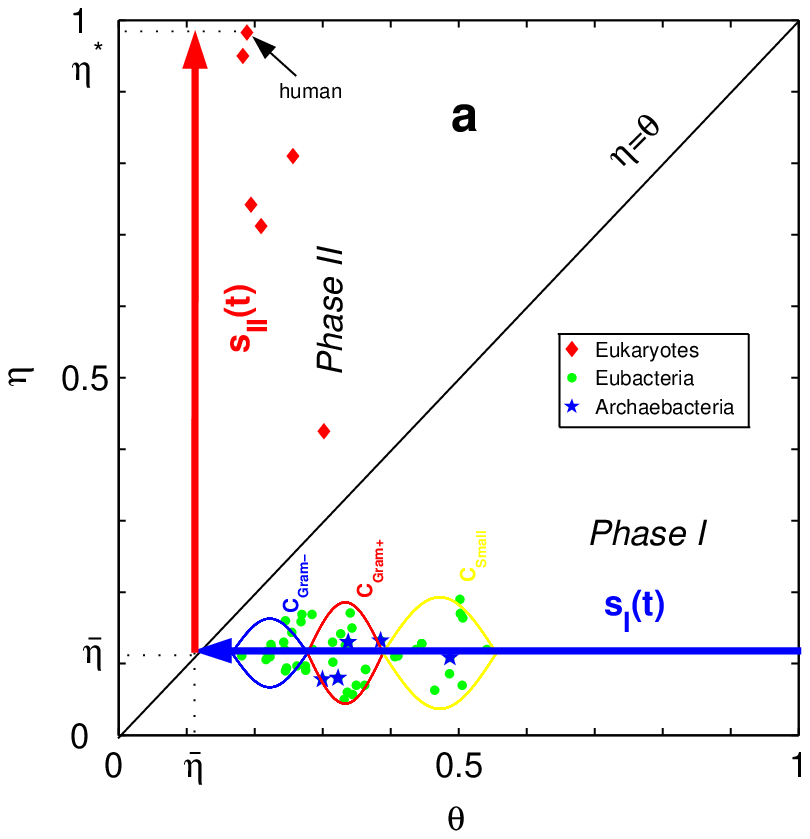}
\includegraphics[width=75mm]{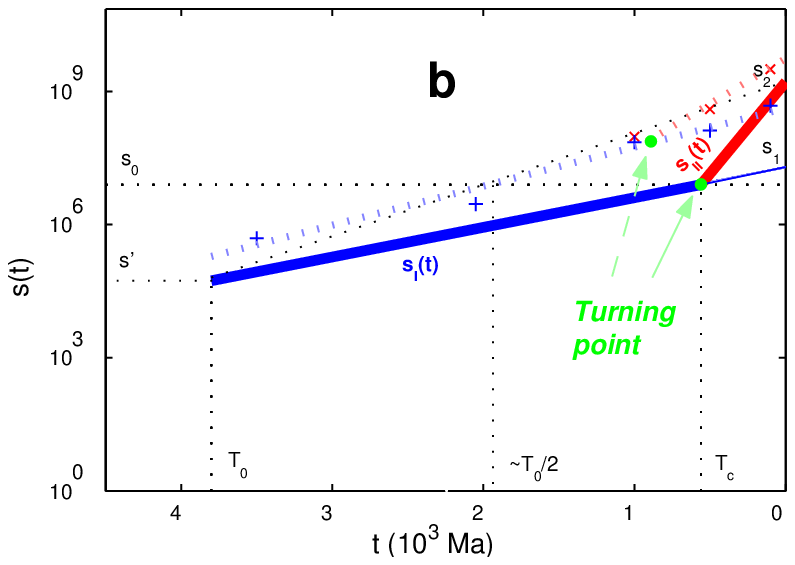}
\includegraphics[width=75mm]{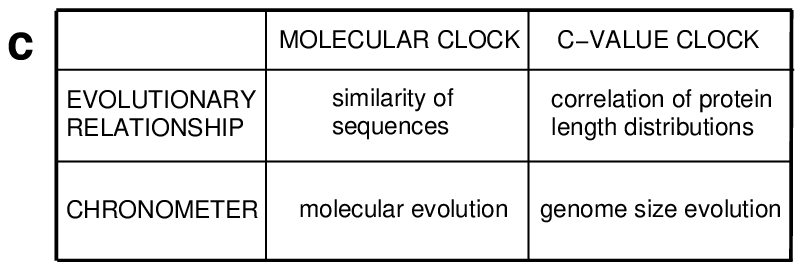}
\includegraphics[width=75mm]{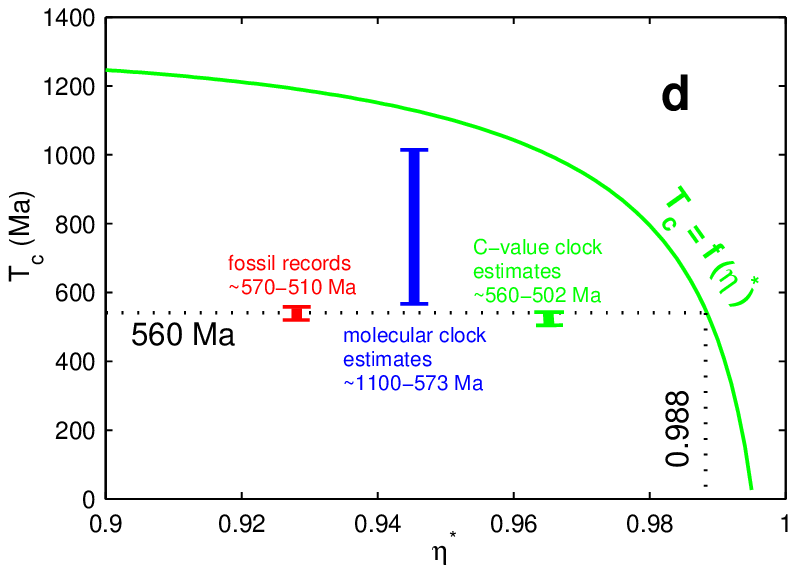}}
\label{fig} \caption{{\bf Genome size evolution and the nature and
timing of the Cambrian explosion.} {\bf a,} The distribution of
species in $\theta-\eta$ plane and the function of genome size
evolution. {\bf b,} The turning point of genome size evolution (red:
total genetic DNA and blue: coding DNA). Our result (solid lines) is
supported by the coarse estimate (thick dotted lines, data for
estimate time and genome size for $5$ taxa are obtained from Ref
\cite{13})). {\bf c,} Comparison between the molecular clock and the
C-value clock. {\bf d,} A sensitive relationship $T_0 = f(\eta^*)$.
If varying $\eta^*$ a little, $T_c$ will change much. The value of
$T_c$ ranges approximately from $502$ Ma to $560$  Ma according to
the C-value clock estimate. The result by C-value clock agrees with
the fossil records \cite{3} better than the molecular clock
estimates \cite{15} \cite{14}. There should be notable systematic
errors in the usual method of molecular clock estimates.}
\end{figure}

\clearpage
\begin{figure}
\centering{
\includegraphics[width=89mm]{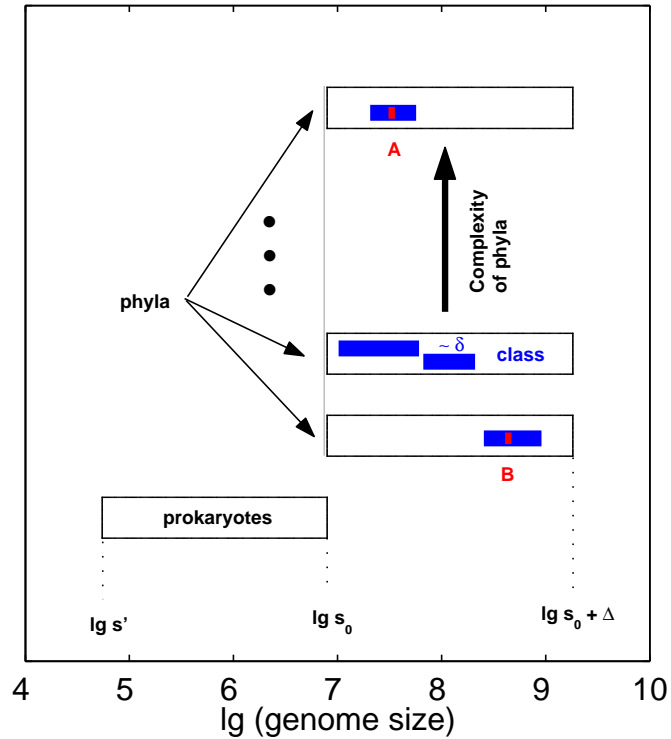}}
\label{fig} \caption{{\bf Explanation of C-value enigma: genome size
range and eukaryotic complexity.} The ranges in genome size by order
of magnitude ($\Delta \sim 2.4$ for phyla and $\delta \sim 0.5$ for
classes) fit the experimental observations in general (see Fig. 1 in
Ref. \cite{28}). In observation, the genome sizes of majority phyla
also vary by about $2$ magnitudes and the genome sizes of majority
classes vary by less than $1$ magnitude \cite{8}. The complexity of
a species inherits from the complexity of the corresponding phylum
in general, so the complexity of species $A$ in a more complex
phylum can potentially outstrip the complexity of species $B$ in a
less complex phylum, though the genome size of $A$ is much less than
that of $B$.}
\end{figure}

\clearpage
\begin{figure}
\centering{
\includegraphics[width=89mm]{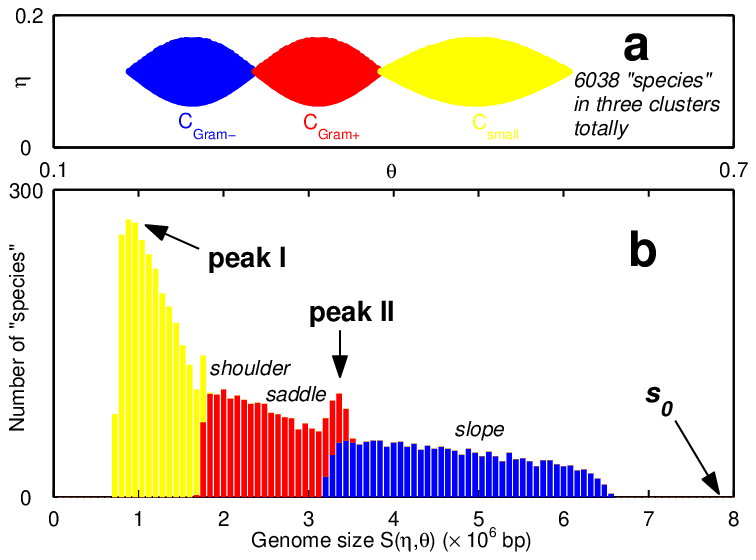}
\includegraphics[width=89mm]{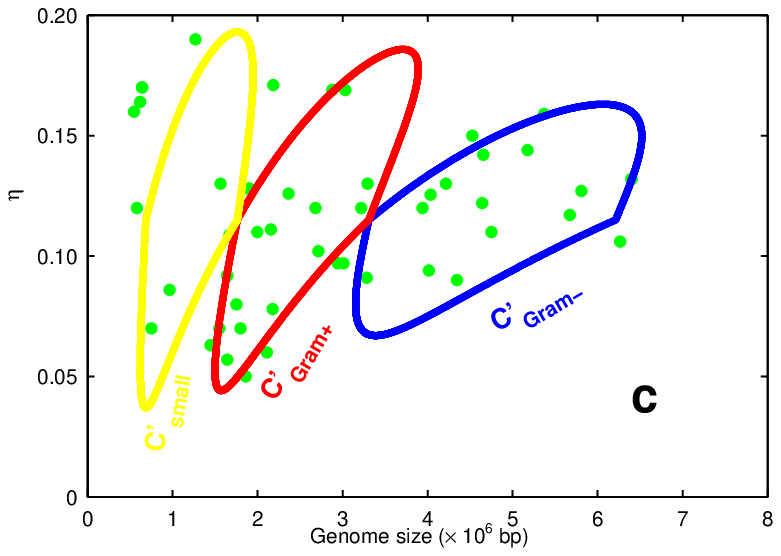}}
\label{fig} \caption{{\bf Explanation of C-value enigma: prokaryotic
genome size distribution.} {\bf a,} Evenly distributed dots
(representing ``species") in three symmetric areas in $\theta-\eta$
plane. {\bf b,} The prediction of prokaryotic genome size
distribution quite fits the experimental observation. The principal
characters of two peaks and their ratio in height and even the
detailed characters such as shoulder, saddle and slope are almost
the same in the actual genome size distribution in Fig. 10.12 in
Ref. \cite{8}. {\bf c,} The prediction of prokaryotic distribution
in $s-\eta$ plane (three asymmetric areas enclosing by lines) quite
fits the intricate distribution of prokaryotes (green dots).}
\end{figure}

\clearpage

\begin{center}
{\Large \bf SUPPLEMENTARY INFORMATION}
\end{center}

\hspace{5cm}
\begin{flushleft}
\hspace{5cm} $\bullet$ Supplementary figures 1 $\sim$ 6 \\
\hspace{5cm} $\bullet$ Supplementary tables 1 $\sim$ 2
\end{flushleft}

\clearpage
%
%

\begin{figure}
\centering{
\includegraphics[width=160mm]{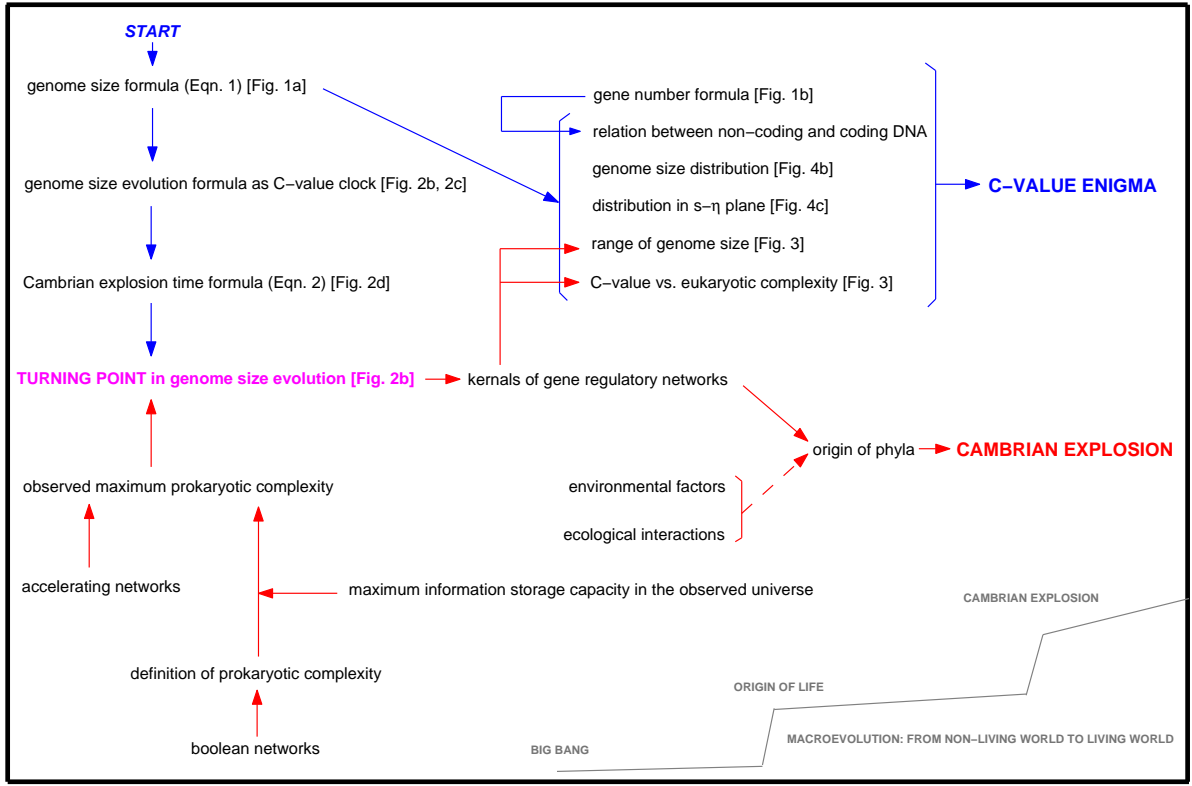}}
\label{fig} \caption{\small {\bf Supplementary Figure 1: The
multidisciplinary framework to explain the Cambrian explosion.} We
found the close relationship between the C-value enigma and the
Cambrian explosion. Hence we invented a new method of C-value clock
depending on the empirical formula of genome size. The unique
turning point in genome size evolution corresponds to the critical
event of the Cambrian explosion. The constraint on the unicellular
genome evolution resulted in the upper limit complexity of
unicellular organisms. We believe that the limited information
storage capacity may determine the complexity of gene networks. The
origin of life and the Cambrian explosion were the most important
milestones in the evolution of biological complexity from nonliving
systems to living systems.}
\end{figure}

\clearpage
\begin{figure}
\centering{
\includegraphics[width=130mm]{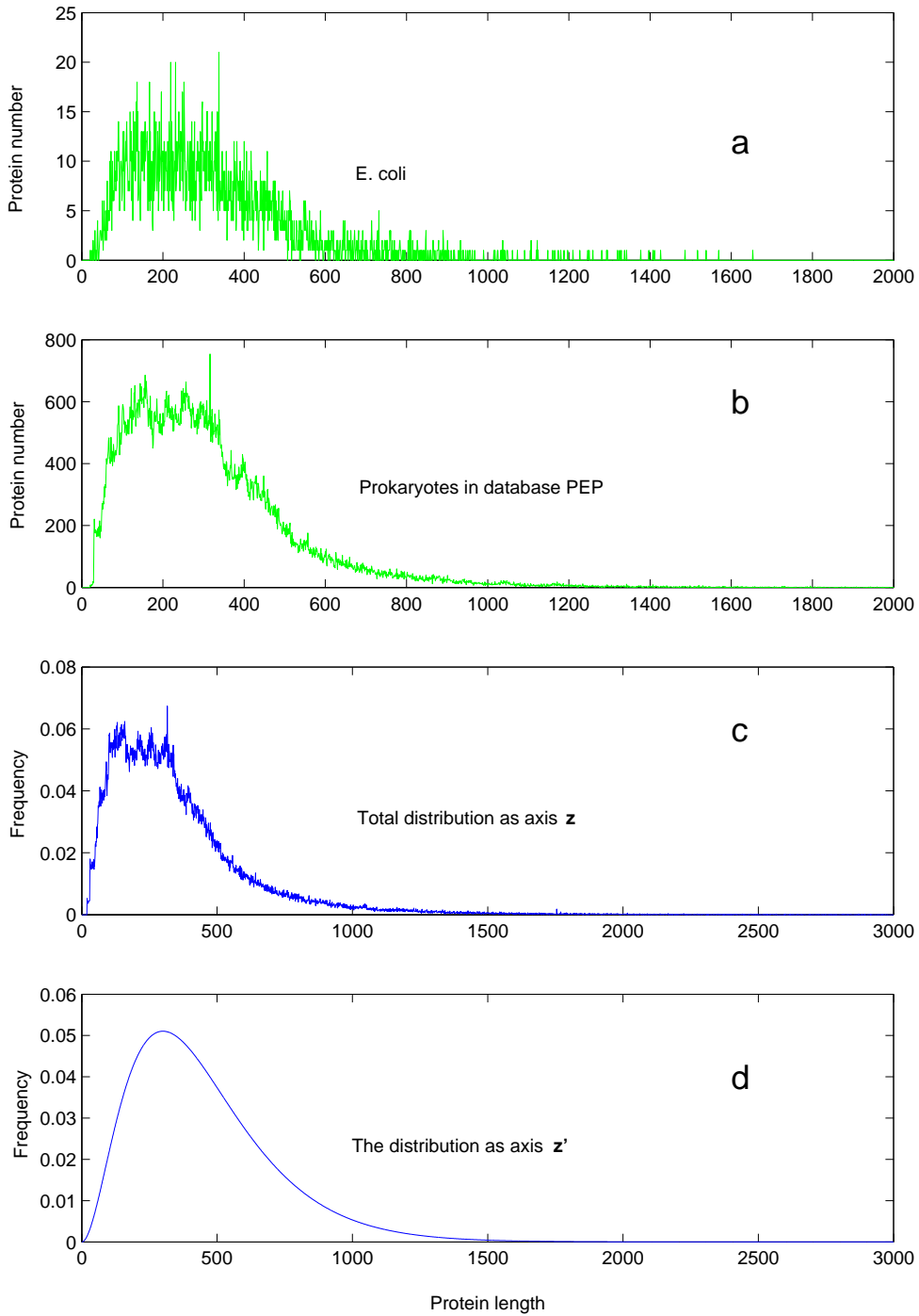}}
\label{fig} \caption{\small {\bf Supplementary Figure 2: Protein
length distributions.} {\bf a}, An example of protein length
distribution of E. coli. {\bf b}, Total protein length distribution
for prokaryotes. {\bf c}, Total protein length distribution for all
the species in database PEP, which can be taken as the polar axis
${\mathbf z}$ in the Supplementary Figure 4a. {\bf d}, An outline of
the protein length distribution, which can be taken as the polar
axis ${\mathbf z}'$ in the Supplementary Figure 4a.}
\end{figure}

\clearpage
\begin{figure}
\centering{
\includegraphics[width=120mm]{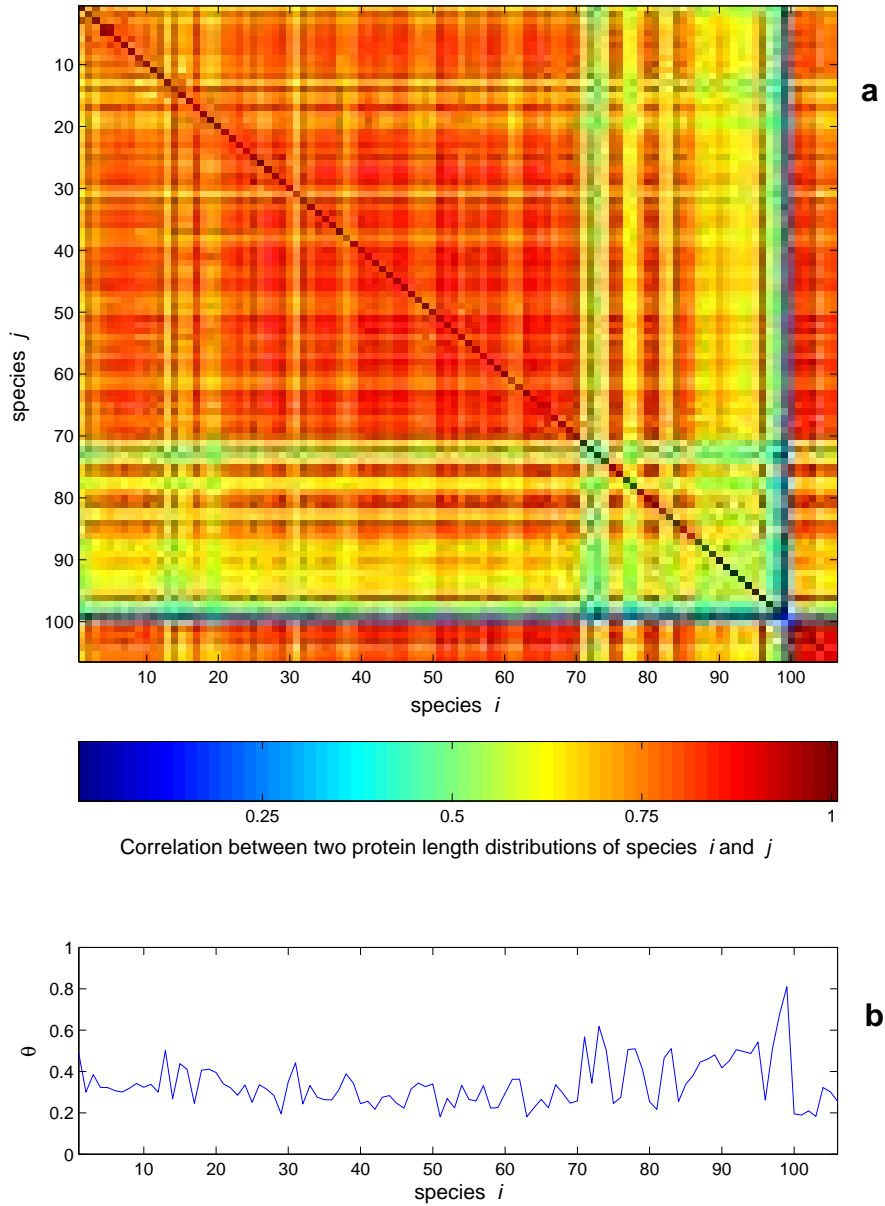}}
\label{fig} \caption{\small {\bf Supplementary Figure 3: The
evolutionary relationship can be revealed by the correlation between
protein length distributions.} {\bf a}, The correlation matrix
$(C_{ij})$ represent the evolutionary relationship between any pairs
of species $i$ and $j$ among the $106$ species. The species in the
matrix are ordered by the average protein length from short to long
for archaebacteria, eubacteria, virus and eukaryotes respectively.
The species can be given concretely by the serial number in
Supplementary Table 1 from the 1st position to the 106th position in
the correlation matrix: 3, 8, 84, 65, 66, 83, 95, 51, 64, 82, 96,
63, 87, 9, 104, 49, 10, 40, 31, 93, 76, 91, 45, 94, 78, 57, 21, 90,
86, 53, 89, 11, 59, 61, 58, 62, 42, 13, 50, 34, 101, 4, 41, 60, 48,
33, 47, 26, 106, 56, 20, 35, 5, 100, 39, 2, 97, 46, 44, 37, 6, 54,
92, 85, 16, 81, 102, 38, 28, 15, 73, 77, 19, 23, 70, 18, 22, 24, 14,
69, 80, 17, 27, 103, 36, 79, 98, 30, 74, 29, 32, 99, 1, 75, 72, 12,
71, 52, 68, 25, 55, 7, 67, 105, 88, 43. {\bf b}, The correlation
polar angle $\theta$  for each of the $106$ species (see
Supplementary Table 2) can be interpreted as the average
evolutionary relationship: the more the average correlation between
protein length distributions is, the less the value of   is; and the
less the value of   is, the closer the average evolutionary
relationship is.}
\end{figure}

\clearpage
\begin{figure}
\centering{
\includegraphics[width=80mm]{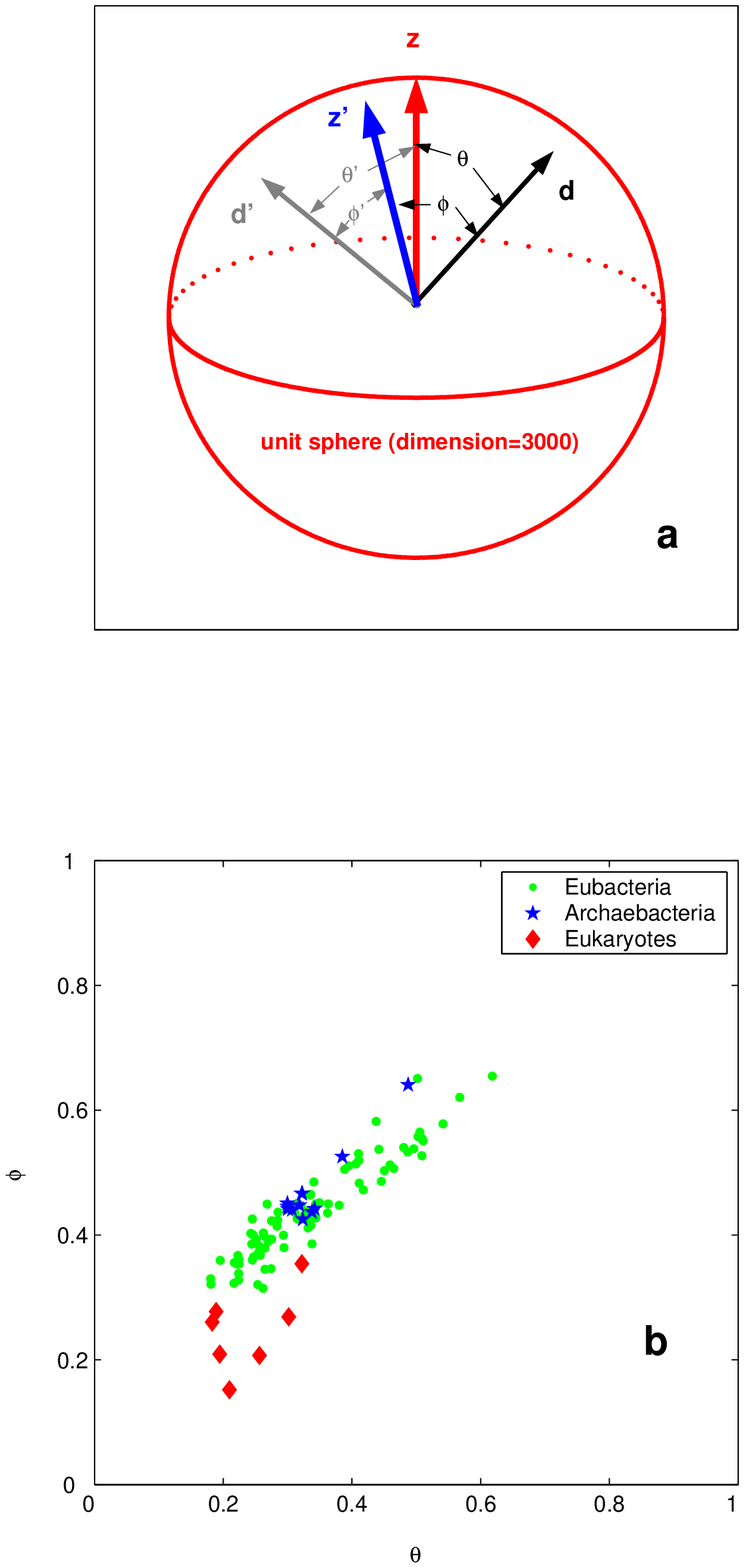}}
\label{fig} \caption{\small {\bf Supplementary Figure 4: The
correlation polar angle and the evolutionary relationship.} {\bf a},
The correlation polar angle $\theta$ and the auxiliary angle $\phi$.
{\bf b}, Distribution of three domains in the $\theta-\eta$ plane.
The evolutionary relationship can be reflected by the correlation
between the protein length distributions. Species in different
domains gather together in different areas respectively.}
\end{figure}

\clearpage
\begin{figure}
\centering{
\includegraphics[width=80mm]{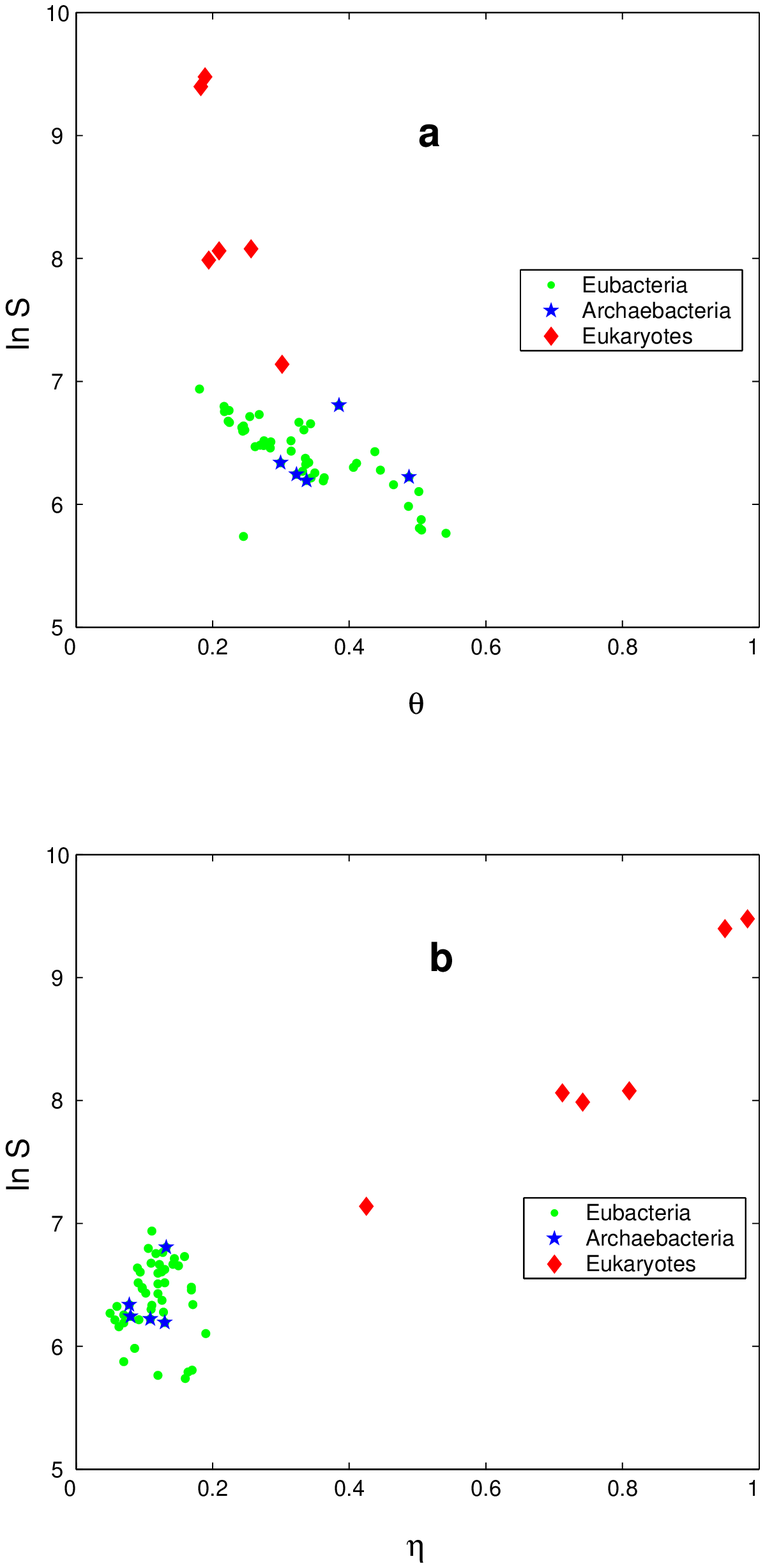}}
\label{fig} \caption{\small {\bf Supplementary Figure 5:
Relationship between the genome size and the non-coding DNA content
and the correlation polar angle.} {\bf a}, $\ln S$increases when
$\theta$ decreases on the whole. {\bf b}, $\ln S$ increases when
$\eta$ increases on the whole.}
\end{figure}

\clearpage
\begin{figure}
\centering{
\includegraphics[width=100mm]{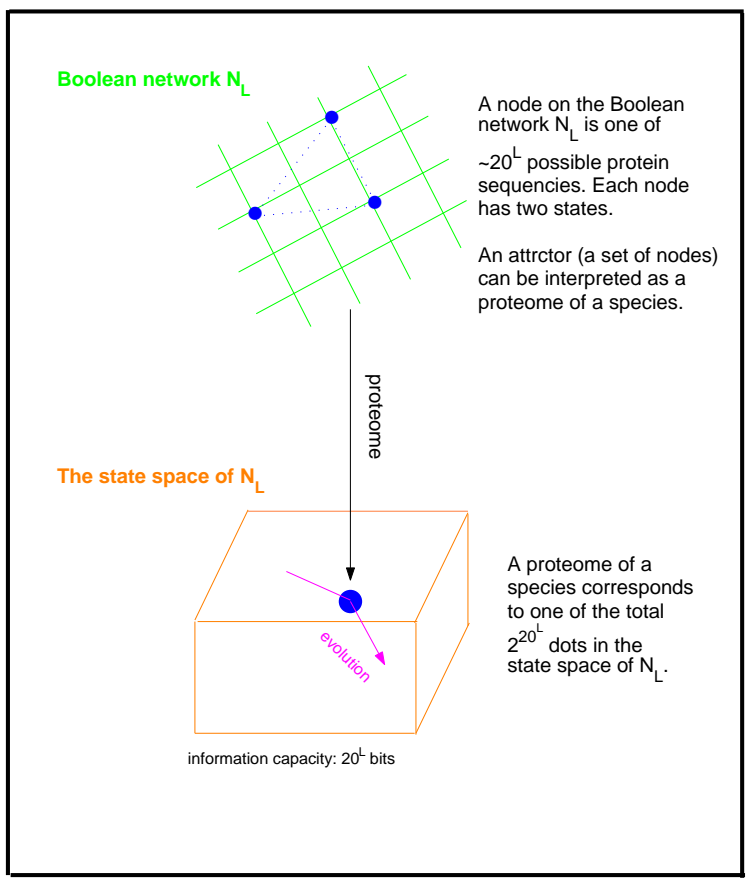}}
\label{fig} \caption{\small {\bf Supplementary Figure 6: Explanation
of prokaryotic complexity by the Boolean network $N_L$  and its
state space.} Each node on the network $N_L$  is one of $\sim 20^L$
possible amino acid sequences, which has two states "on" or "off"
according to the theory of Boolean networks. Each point in the state
space of $N_L$ represents a "proteome" (a set of "proteins" as an
attractor of the Boolean network  $N_L$ whose states are "on"). The
attractor is robust against the perturbations in evolution. The
evolution of a species can be described by a trajectory of the
evolving proteome of the species in the state space of $N_L$. An
underlying evolutionary mechanism is necessary to determine the
movement of the species in the global state space of $N_L$, so the
complexity of the life system is proportional to the number of
points in the state space of $N_L$. The information stored in gene
networks (~ $20^L$ bits) reflects the complexity of the life system,
which is compatible to the maximum information stored in the
observed universe $I_{univ}$.}
\end{figure}

\clearpage

\begin{center}{\Large {\bf Supplementary Table 1:}
Organisms in the database Predictions for Entire Proteomes £¨PEP£©}
\end{center}

\vspace{1cm} Notes: There are 7 eukaryotes, 12 archaebecteria, 85
eubacteria and 2 viruses in PEP.

\newpage

\begin{tabular}{|l|}

\hline
(No. 1)  \\
PEP FILE:   achfl.pep\\
ORGANISM:   Acholeplasma florum (Mesoplasma florum); A (M) florum; achfl\\
DOMAIN:     Eubacteria\\
\hline
(No. 2)  \\
PEP FILE:   aciad.pep\\
ORGANISM:   Acinetobacter sp (strain ADP1); A sp ADP1; aciad\\
DOMAIN:     Eubacteria\\
\hline
(No. 3)  \\
PEP FILE:   aerpe.pep\\
ORGANISM:   Aeropyrum pernix K1; A pernix K1; aerpe\\
DOMAIN:     Archaebacteria\\
\hline
(No. 4)  \\
PEP FILE:   agrt5.pep\\
ORGANISM:   Agrobacterium tumefaciens (strain C58 / ATCC 33970); A tumefaciens; agrt5\\
DOMAIN:     Eubacteria\\
\hline
(No. 5)  \\
PEP FILE:   agrtu.pep\\
ORGANISM:   Agrobacterium tumefaciens; A tumefaciens; agrtu\\
DOMAIN:     Eubacteria\\
\hline
(No. 6)  \\
PEP FILE:   aquae.pep\\
ORGANISM:   Aquifex aeolicus; A aeolicus; aquae\\
DOMAIN:     Eubacteria\\
\hline
(No. 7)  \\
PEP FILE:   arath.pep\\
ORGANISM:   Arabidopsis thaliana; A thaliana; arath\\
DOMAIN:     Eukaryote\\
\hline
(No. 8)  \\
PEP FILE:   arcfu.pep\\
ORGANISM:   Achaeoglobus fulgidus; A fulgidus; arcfu\\
DOMAIN:     Archaebacteria\\
\hline
(No. 9)  \\
PEP FILE:   bacaa.pep\\
ORGANISM:   Bacillus anthracis (strain Ames); B anthracis\_Ames; bacaa\\
DOMAIN:     Eubacteria\\
\hline
(No. 10)  \\
PEP FILE:   bacce.pep\\
ORGANISM:   Bacillus cereus (ATCC 14579); B cereus (ATCC 14579); bacce\\
DOMAIN:     Eubacteria\\
\hline

\end{tabular}
\clearpage
\begin{tabular}{|l|}
\hline

(No. 11)  \\
PEP FILE:   bacsu.pep\\
ORGANISM:   Bacillus subtilis; B subtilis; bacsu\\
DOMAIN:     Eubacteria\\
\hline
(No. 12)  \\
PEP FILE:   bactn.pep\\
ORGANISM:   Bacteroides thetaiotaomicron VPI-5482; B thetaiotaomicron VPI-5482; bactn\\
DOMAIN:     Eubacteria\\
\hline
(No. 13)  \\
PEP FILE:   barhe.pep\\
ORGANISM:   Bartonella henselae (Houston-1); B henselae Houston-1; barhe\\
DOMAIN:     Eubacteria\\
\hline
(No. 14)  \\
PEP FILE:   barqu.pep\\
ORGANISM:   Bartonella quintana (Toulouse); B quintana Toulouse; barqu\\
DOMAIN:     Eubacteria\\
\hline
(No. 15)  \\
PEP FILE:   bdeba.pep\\
ORGANISM:   Bdellovibrio bacteriovorus; B bacteriovorus; bdeba\\
DOMAIN:     Eubacteria\\
\hline
(No. 16)  \\
PEP FILE:   borbr.pep\\
ORGANISM:   Bordetella bronchiseptica RB50; B bronchiseptica RB50; borbr\\
DOMAIN:     Eubacteria\\
\hline
(No. 17)  \\
PEP FILE:   borbu.pep\\
ORGANISM:   Borrelia burgdorferi; B burgdorferi; borbu\\
DOMAIN:     Eubacteria\\
\hline
(No. 18)  \\
PEP FILE:   borpa.pep\\
ORGANISM:   Bordetella parapertussis; B parapertussis; borpa\\
DOMAIN:     Eubacteria\\
\hline
(No. 19)  \\
PEP FILE:   borpe.pep\\
ORGANISM:   Bordetella pertussis; B pertussis; borpe\\
DOMAIN:     Eubacteria\\
\hline
(No. 20)  \\
PEP FILE:   braja.pep\\
ORGANISM:   Bradyrhizobium japonicum; B japonicum; braja\\
DOMAIN:     Eubacteria\\
\hline

\end{tabular}
\newpage
\begin{tabular}{|l|}
\hline

(No. 21)  \\
PEP FILE:   brume.pep\\
ORGANISM:   Brucella melitensis; B melitensis; brume\\
DOMAIN:     Eubacteria\\
\hline
(No. 22)  \\
PEP FILE:   bucai.pep\\
ORGANISM:   Buchnera aphidicola (subsp. Acyrthosiphon pisum); B aphidicola (subsp.\\ \vspace{-0.1cm}Acyrthosiphon pisum); bucai\\
DOMAIN:     Eubacteria\\
\hline
(No. 23)  \\
PEP FILE:   bucap.pep\\
ORGANISM:   Buchnera aphidicola (subsp. Schizaphis graminum); B aphidicola (subsp.\\ \vspace{-0.1cm}Schizaphis graminum); bucap\\
DOMAIN:     Eubacteria\\
\hline
(No. 24)  \\
PEP FILE:   bucbp.pep\\
ORGANISM:   Buchnera aphidicola (subsp. Baizongia pistaciae); B aphidicola (subsp.\\ \vspace{-0.1cm}Baizongia pistaciae); bucbp\\
DOMAIN:     Eubacteria\\
\hline
(No. 25)  \\
PEP FILE:   caeel.pep\\
ORGANISM:   Caenorhabditis elegans; C elegans; caeel\\
DOMAIN:     Eukaryote\\
\hline
(No. 26)  \\
PEP FILE:   camje.pep\\
ORGANISM:   Campylobacter jejuni; C jejuni; camje\\
DOMAIN:     Eubacteria\\
\hline
(No. 27)  \\
PEP FILE:   canbf.pep\\
ORGANISM:   Candidatus Blochmannia floridanus; C Blochmannia floridanus; canbf\\
DOMAIN:     Eubacteria\\
\hline
(No. 28)  \\
PEP FILE:   caucr.pep\\
ORGANISM:   Caulobacter crescentus; C crescentus; caucr\\
DOMAIN:     Eubacteria\\
\hline
(No. 29)  \\
PEP FILE:   chlcv.pep\\
ORGANISM:   Chlamydophila caviae; C caviae; chlcv\\
DOMAIN:     Eubacteria\\
\hline
(No. 30)  \\
PEP FILE:   chlmu.pep\\
ORGANISM:   Chlamydia muridarum; C muridarum; chlmu\\
DOMAIN:     Eubacteria\\
\hline

\end{tabular}
\clearpage
\begin{tabular}{|l|}
\hline

(No. 31)  \\
PEP FILE:   chlte.pep\\
ORGANISM:   Chlorobium tepidum; C tepidum; chlte\\
DOMAIN:     Eubacteria\\
\hline
(No. 32)  \\
PEP FILE:   chltr.pep\\
ORGANISM:   Chlamydia trachomatis; C trachomatis; chltr\\
DOMAIN:     Eubacteria\\
\hline
(No. 33)  \\
PEP FILE:   chrvo.pep\\
ORGANISM:   Chromobacterium violaceum ATCC 12472; C violaceum ATCC 12472; chrvo\\
DOMAIN:     Eubacteria\\
\hline
(No. 34)  \\
PEP FILE:   cloab.pep\\
ORGANISM:   Clostridium acetobutylicum; C acetobutylicum; cloab\\
DOMAIN:     Eubacteria\\
\hline
(No. 35)  \\
PEP FILE:   clope.pep\\
ORGANISM:   Clostridium perfringens; C perfringens; clope\\
DOMAIN:     Eubacteria\\
\hline
(No. 36)  \\
PEP FILE:   clote.pep\\
ORGANISM:   Clostridium tetani; C tetani; clote\\
DOMAIN:     Eubacteria\\
\hline
(No. 37)  \\
PEP FILE:   cordi.pep\\
ORGANISM:   Corynebacterium diphtheriae NCTC 13129; C diphtheriae NCTC 13129; cordi\\
DOMAIN:     Eubacteria\\
\hline
(No. 38)  \\
PEP FILE:   coref.pep\\
ORGANISM:   Corynebacterium efficiens; C efficiens; coref\\
DOMAIN:     Eubacteria\\
\hline
(No. 39)  \\
PEP FILE:   corgl.pep\\
ORGANISM:   Corynebacterium glutamicum; C glutamicum; corgl\\
DOMAIN:     Eubacteria\\
\hline
(No. 40)  \\
PEP FILE:   coxbu.pep\\
ORGANISM:   Coxiella burnetii; C burnetii; coxbu\\
DOMAIN:     Eubacteria\\
\hline

\end{tabular}
\clearpage
\begin{tabular}{|l|}
\hline

(No. 41)  \\
PEP FILE:   deira.pep  \hspace{12cm}\ \\
ORGANISM:   Deinococcus radiodurans; D radiodurans; deira\\
DOMAIN:     Eubacteria\\
\hline
(No. 42)  \\
PEP FILE:   desvh.pep\\
ORGANISM:   Desulfovibrio vulgaris subsp. vulgaris str. Hildenborough;\\ D vulgaris subsp. vulgaris str. Hildenborough; desvh\\
DOMAIN:     Eubacteria\\
\hline
(No. 43)  \\
PEP FILE:   drome.pep\\
ORGANISM:   Drosophila melanogaster; D melanogaster; drome\\
DOMAIN:     Eukaryote\\
\hline
(No. 44)  \\
PEP FILE:   ecoli.pep\\
ORGANISM:   Escherichia coli; E coli; ecoli\\
DOMAIN:     Eubacteria\\
\hline
(No. 45)  \\
PEP FILE:   entfa.pep\\
ORGANISM:   Enterococcus faecalis; E faecalis; entfa\\
DOMAIN:     Eubacteria\\
\hline
(No. 46)  \\
PEP FILE:   erwca.pep\\
ORGANISM:   Erwinia carotovora; E carotovora; erwca\\
DOMAIN:     Eubacteria\\
\hline
(No. 47)  \\
PEP FILE:   fusnu.pep\\
ORGANISM:   Fusobacterium nucleatum; F nucleatum; fusnu\\
DOMAIN:     Eubacteria\\
\hline
(No. 48)  \\
PEP FILE:   glovi.pep\\
ORGANISM:   Gloeobacter violaceus; G violaceus; glovi\\
DOMAIN:     Eubacteria\\
\hline
(No. 49)  \\
PEP FILE:   haedu.pep\\
ORGANISM:   Haemophilus ducreyi; H ducreyi; haedu\\
DOMAIN:     Eubacteria\\
\hline
(No. 50)  \\
PEP FILE:   haein.pep\\
ORGANISM:   Haemophilus influenzae; H influenzae; haein\\
DOMAIN:     Eubacteria\\
\hline

\end{tabular}
\clearpage
\begin{tabular}{|l|}
\hline

(No. 51)  \\
PEP FILE:   haln1.pep\\
ORGANISM:   Halobacterium sp. (strain NRC-1); H sp. (strain NRC-1); haln1\\
DOMAIN:     Archaebacteria\\
\hline
(No. 52)  \\
PEP FILE:   hcmva.pep\\
ORGANISM:   Human cytomegalovirus (strain AD169); HCMV (strain AD169); hcmva\\
DOMAIN:     virus\\
\hline
(No. 53)  \\
PEP FILE:   helhe.pep\\
ORGANISM:   Helicobacter heilmannii; H heilmannii; helhe\\
DOMAIN:     Eubacteria\\
\hline
(No. 54)  \\
PEP FILE:   helpy.pep\\
ORGANISM:   Helicobacter pylori; H pylori; helpy\\
DOMAIN:     Eubacteria\\
\hline
(No. 55)\\
PEP FILE:   human.pep\\
ORGANISM:   Homo sapiens; H sapiens; human\\
DOMAIN:     Eukaryote\\
\hline
(No. 56)  \\
PEP FILE:   lacjo.pep\\
ORGANISM:   Lactobacillus johnsonii; L johnsonii; lacjo\\
DOMAIN:     Eubacteria\\
\hline
(No. 57)  \\
PEP FILE:   lacla.pep\\
ORGANISM:   Lactococcus lactis (subsp. lactis); L lactis (subsp. lactis); lacla\\
DOMAIN:     Eubacteria\\
\hline
(No. 58)  \\
PEP FILE:   lacpl.pep\\
ORGANISM:   Lactobacillus plantarum WCFS1; L plantarum WCFS1; lacpl\\
DOMAIN:     Eubacteria\\
\hline
(No. 59)  \\
PEP FILE:   leixx.pep\\
ORGANISM:   Leifsonia xyli (subsp. xyli); L xyli (subsp. xyli); leixx\\
DOMAIN:     Eubacteria\\
\hline
(No. 60)  \\
PEP FILE:   lepic.pep\\
ORGANISM:   Leptospira interrogans (serogroup Icterohaemorrhagiae / serovar Copenhageni);\\ L interrogans (serogroup Icterohaemorrhagiae / serovar Copenhageni); lepic\\
DOMAIN:     Eubacteria\\
\hline

\end{tabular}
\clearpage
\begin{tabular}{|l|}
\hline

(No. 61)  \\
PEP FILE:   lisin.pep\\
ORGANISM:   Listeria innocua; L innocua; lisin\\
DOMAIN:     Eubacteria\\
\hline
(No. 62)  \\
PEP FILE:   lismo.pep\\
ORGANISM:   Listeria monocytogenes; L monocytogenes; lismo\\
DOMAIN:     Eubacteria\\
\hline
(No. 63)  \\
PEP FILE:   metac.pep\\
ORGANISM:   Methanosarcina acetivorans; M acetivorans; metac\\
DOMAIN:     Archaebacteria\\
\hline
(No. 64)  \\
PEP FILE:   metka.pep\\
ORGANISM:   Methanopyrus kandleri; M kandleri; metka\\
DOMAIN:     Archaebacteria\\
\hline
(No. 65)  \\
PEP FILE:   metth.pep\\
ORGANISM:   Methanobacterium thermoautotrophicum; M thermoautotrophicum; metth\\
DOMAIN:     Archaebacteria\\
\hline
(No. 66)  \\
PEP FILE:   mettm.pep\\
ORGANISM:   Methanobacterium thermoautotrophicum; M thermoautotrophicum ; mettm\\
DOMAIN:     Archaebacteria\\
\hline
(No. 67) \\
PEP FILE:   mouse.pep\\
ORGANISM:   Mus musculus; M musculus; mouse\\
DOMAIN:     Eukaryote\\
\hline
(No. 68)  \\
PEP FILE:   muhv4.pep\\
ORGANISM:   Murine herpesvirus 68 strain WUMS; Murine herpesvirus 68 strain WUMS;\\ muhv4\\
DOMAIN:     virus\\
\hline
(No. 69)  \\
PEP FILE:   mycav.pep\\
ORGANISM:   Mycobacterium avium; M avium; mycav\\
DOMAIN:     Eubacteria\\
\hline
(No. 70)  \\
PEP FILE:   mycbo.pep\\
ORGANISM:   Mycobacterium bovis AF2122/97; M bovis AF2122/97; mycbo\\
DOMAIN:     Eubacteria\\
\hline

\end{tabular}
\clearpage
\begin{tabular}{|l|}
\hline

(No. 71)  \\
PEP FILE:   mycga.pep \\
ORGANISM:   Mycoplasma gallisepticum; M gallisepticum; mycga\\
DOMAIN:     Eubacteria\\
\hline
(No. 72)  \\
PEP FILE:   mycge.pep\\
ORGANISM:   Mycoplasma genitalium; M genitalium; mycge\\
DOMAIN:     Eubacteria\\
\hline
(No. 73)  \\
PEP FILE:   mycms.pep\\
ORGANISM:   Mycoplasma mycoides (subsp. mycoides SC); M mycoides (subsp. mycoides\\ SC); mycms\\
DOMAIN:     Eubacteria\\
\hline
(No. 74)  \\
PEP FILE:   mycpn.pep\\
ORGANISM:   Mycoplasma pneumoniae; M pneumoniae; mycpn\\
DOMAIN:     Eubacteria\\
\hline
(No. 75)  \\
PEP FILE:   mycpu.pep\\
ORGANISM:   Mycoplasma pulmonis; M pulmonis; mycpu\\
DOMAIN:     Eubacteria\\
\hline
(No. 76)  \\
PEP FILE:   neime.pep\\
ORGANISM:   Neisseria meningitidis; N meningitidis; neime\\
DOMAIN:     Eubacteria\\
\hline
(No. 77)  \\
PEP FILE:   niteu.pep\\
ORGANISM:   Nitrosomonas europaea; N europaea; niteu\\
DOMAIN:     Eubacteria\\
\hline
(No. 78)  \\
PEP FILE:   oceih.pep\\
ORGANISM:   Oceanobacillus iheyensis; O iheyensis; oceih\\
DOMAIN:     Eubacteria\\
\hline
(No. 79)  \\
PEP FILE:   porgi.pep\\
ORGANISM:   Porphyromonas gingivalis; P gingivalis; porgi\\
DOMAIN:     Eubacteria\\
\hline
(No. 80)  \\
PEP FILE:   pseae.pep\\
ORGANISM:   Pseudomonas aeruginosa; P aeruginosa; pseae\\
DOMAIN:     Eubacteria\\
\hline

\end{tabular}
\clearpage
\begin{tabular}{|l|}
\hline

(No. 81)  \\
PEP FILE:   psepu.pep  \hspace{12cm}\ \\
ORGANISM:   Pseudomonas putida; P putida; psepu\\
DOMAIN:     Eubacteria\\
\hline
(No. 82)  \\
PEP FILE:   pyrab.pep\\
ORGANISM:   Pyrococcus abyssi; P abyssi; pyrab\\
DOMAIN:     Archaebacteria\\
\hline
(No. 83)  \\
PEP FILE:   pyrfu.pep\\
ORGANISM:   Pyrococcus furiosus; P furiosus; pyrfu\\
DOMAIN:     Archaebacteria\\
\hline
(No. 84)  \\
PEP FILE:   pyrho.pep\\
ORGANISM:   Pyrococcus horikoshii; P horikoshii; pyrho\\
DOMAIN:     Archaebacteria\\
\hline
(No. 85)  \\
PEP FILE:   ralso.pep\\
ORGANISM:   Ralstonia solanacearum; R solanacearum; ralso\\
DOMAIN:     Eubacteria\\
\hline
(No. 86)  \\
PEP FILE:   rhilo.pep\\
ORGANISM:   Rhizobium loti; R loti; rhilo\\
DOMAIN:     Eubacteria\\
\hline
(No. 87)  \\
PEP FILE:   riccn.pep\\
ORGANISM:   Rickettsia conorii; R conorii; riccn\\
DOMAIN:     Eubacteria\\
\hline
(No. 88)  \\
PEP FILE:   schpo.pep
\\
PEP FILE:   SPBC839\_05c
ORG Schizosaccharomyces pombe; S pombe; schpo\\
DOMAIN:     Eukaryote\\
\hline
(No. 89)  \\
PEP FILE:   shifl.pep\\
ORGANISM:   Shigella flexneri; S flexneri; shifl\\
DOMAIN:     Eubacteria\\
\hline
(No. 90)  \\
PEP FILE:   staau.pep\\
ORGANISM:   Staphylococcus aureus; S aureus; staau\\
DOMAIN:     Eubacteria\\
\hline

\end{tabular}
\clearpage
\begin{tabular}{|l|}
\hline

(No. 91)  \\
PEP FILE:   strag.pep  \hspace{12cm}\ \\
ORGANISM:   Streptococcus agalactiae; S agalactiae; strag\\
DOMAIN:     Eubacteria\\
\hline
(No. 92)  \\
PEP FILE:   strco.pep\\
ORGANISM:   Streptomyces coelicolor; S coelicolor; strco\\
DOMAIN:     Eubacteria\\
\hline
(No. 93)  \\
PEP FILE:   strpn.pep\\
ORGANISM:   Streptococcus pneumoniae; S pneumoniae; strpn\\
DOMAIN:     Eubacteria\\
\hline
(No. 94)  \\
PEP FILE:   strpy.pep\\
ORGANISM:   Streptococcus pyogenes; S pyogenes; strpy\\
DOMAIN:     Eubacteria\\
\hline
(No. 95)  \\
PEP FILE:   sulso.pep\\
ORGANISM:   Sulfolobus solfataricus; S solfataricus; sulso\\
DOMAIN:     Archaebacteria\\
\hline
(No. 96)  \\
PEP FILE:   theac.pep\\
ORGANISM:   Thermoplasma acidophilum; T acidophilum; theac\\
DOMAIN:     Archaebacteria\\
\hline
(No. 97)  \\
PEP FILE:   thema.pep\\
ORGANISM:   Thermotoga maritima; T maritima; thema\\
DOMAIN:     Eubacteria\\
\hline
(No. 98)  \\
PEP FILE:   trepa.pep\\
ORGANISM:   Treponema pallidum; T pallidum; trepa\\
DOMAIN:     Eubacteria\\
\hline
(No. 99)  \\
PEP FILE:   ureur.pep\\
ORGANISM:   Ureaplasma urealyticum; U urealyticum; ureur\\
DOMAIN:     Eubacteria\\
\hline
(No. 100)  \\
PEP FILE:   vibch.pep\\
ORGANISM:   Vibrio cholerae; V cholerae; vibch\\
DOMAIN:     Eubacteria\\
\hline

\end{tabular}
\clearpage
\begin{tabular}{|l|}
\hline

(No. 101)  \\
PEP FILE:   vibpa.pep\\
ORGANISM:   Vibrio parahaemolyticus RIMD 2210633; V parahaemolyticus RIMD 2210633;\\ vibpa\\
DOMAIN:     Eubacteria\\
\hline
(No. 102)  \\
PEP FILE:   wolsu.pep\\
ORGANISM:   Wolinella succinogenes; W succinogenes; wolsu\\
DOMAIN:     Eubacteria\\
\hline
(No. 103)  \\
PEP FILE:   xanac.pep\\
ORGANISM:   Xanthomonas axonopodis (pv. citri); X axonopodis (pv. citri); xanac\\
DOMAIN:     Eubacteria\\
\hline
(No. 104)  \\
PEP FILE:   xylfa.pep\\
ORGANISM:   Xylella fastidiosa; X fastidiosa; xylfa\\
DOMAIN:     Eubacteria\\
\hline
(No. 105)  \\
PEP FILE:   yeast.pep\\
ORGANISM:   Saccharomyces cerevisiae; S cerevisiae; yeast\\
DOMAIN:     Eukaryote\\
\hline
(No. 106)  \\
PEP FILE:   yerpe.pep\\
ORGANISM:   Yersinia pestis; Y pestis; yerpe\\
DOMAIN:     Eubacteria\\

\hline

\end{tabular}

\clearpage

\begin{center}{\Large {\bf Supplementary Table 2:}
Data of $\eta$, $\theta$ and the comparison between theoretical
predictions and experimental observations for genome size and gene
number.}
\end{center}

\vspace{1cm} Notes: The serial numbers for organisms here are the
same numbers for the organisms in Supplementary Table 1. The data of
non-coding DNA contents $\eta$ and the genome sizes are obtained
from Ref. \cite{17}, where there are $\mathbf 54$ species ($\mathbf
6$ eukaryotes, $\mathbf 5$ archaebacteria and $\mathbf 43$
eubacteria, i.e., $\mathbf 48$ prokaryotes in total) can be also
found in database PEP. The gene numbers are obtained by the numbers
of Open Reading Frames (ORFs) in proteomes in PEP. The non-coding
content is obtained according to the Human genome draft in this
table according to Ref. \cite{17}. But we choose the more precise
value of $\eta^*$ according to the finished euchromatic sequence of
the human genome in Ref. \cite{16} to calculate the accurate time of
the Cambrian explosion.

\newpage

\begin{tabular}{|c|c|c|c|c|c|c|}

\hline

No. & $\eta$  &  $\theta$  &  genome size &  $S(\eta,\theta)$ & gene
number  & $N(\eta, \theta)$

\\ \hline

 1 & & 0.4960 & & & 683 & \\ \hline
 2 & & 0.2571 & & & 3322 & \\ \hline
 3 & 0.1088 & 0.4874 & 1669695 & 9.6490e+005 & 2694 & 839.3707 \\ \hline
 4 & 0.1170 & 0.2173 & 5674062 & 4.7051e+006 & 5402 & 4.7732e+003 \\ \hline
 5 & & 0.2238 & & & 5274 & \\ \hline
 6 & 0.0700 & 0.3620 & 1551335 & 1.5559e+006 & 1522 & 1.7165e+003 \\ \hline
 7 & 0.7120 & 0.2096 & 115409949 & 1.8103e+008 & 25541 & 1.8126e+004 \\ \hline
 8 & 0.0780 & 0.2996 & 2178400 & 2.3277e+006 & 2406 & 2.5982e+003 \\ \hline
 9 & 0.1590 & 0.2681 & 5370060 & 4.5484e+006 & 5311 & 3.7826e+003 \\ \hline
 10 & 0.1600 & 0.2452 & 546909 & 5.2118e+006 & 5274 & 4.3859e+003 \\ \hline
 11 & 0.1300 & 0.2428 & 4214810 & 4.4046e+006 & 4099 & 4.1737e+003 \\ \hline
 12 & & 0.2617 & & & 4776 & \\ \hline
 13 & & 0.3886 & & & 1482 & \\ \hline
 14 & & 0.4112 & & & 1141 & \\ \hline
 15 & & 0.2575 & & & 3584 & \\ \hline
 16 & & 0.2649 & & & 4986 & \\ \hline
 17 & 0.0630 & 0.4649 & 1443725 & 8.3102e+005 & 850 & 877.8284 \\ \hline
 18 & & 0.2744 & & & 4184 & \\ \hline
 19 & & 0.6183 & & & 3446 & \\ \hline
 20 & & 0.1805 & & & 8307 & \\ \hline
 21 & 0.1300 & 0.3146 & 3294935 & 2.9287e+006 & 2059 & 2.6414e+003 \\ \hline
 22 & 0.1640 & 0.5060 & 618000 & 1.2136e+006 & 574 & 840.3921 \\ \hline
 23 & 0.1700 & 0.5028 & 640000 & 1.2814e+006 & 546 & 868.7162 \\ \hline
 24 & & 0.5092 & & & 504 & \\ \hline
 25 & 0.7419 & 0.1945 & 97000000 & 2.3647e+008 & 21832 & 2.1291e+004 \\ \hline
 26 & 0.0570 & 0.3441 & 1641181 & 1.5915e+006 & 1633 & 1.8700e+003 \\ \hline
 27 & & 0.5105 & & & 583 & \\ \hline
 28 & 0.0940 & 0.2471 & 4016942 & 3.4556e+006 & 3737 & 3.7569e+003 \\ \hline
 29 & & 0.4177 & & & 998 & \\ \hline
 30 & & 0.4590 & & & 907 & \\ \hline
 31 & 0.1110 & 0.4107 & 2154946 & 1.5126e+006 & 2252 & 1.3753e+003 \\ \hline
 32 & & 0.4502 & & & 894 & \\ \hline
 33 & 0.1100 & 0.2228 & 4751080 & 4.3729e+006 & 4396 & 4.5420e+003 \\ \hline
 34 & 0.1200 & 0.2442 & 3940880 & 4.1134e+006 & 3847 & 4.0489e+003 \\ \hline
 35 & 0.1690 & 0.2695 & 3031430 & 4.7935e+006 & 2722 & 3.8301e+003 \\ \hline
 36 & & 0.3379 & & & 2373 & \\ \hline
 37 & & 0.2937 & & & 2269 & \\ \hline
 38 & & 0.2943 & & & 2947 & \\ \hline
 39 & & 0.2645 & & & 2989 & \\ \hline
 40 & 0.1100 & 0.4060 & 1995275 & 1.5434e+006 & 2009 & 1.4133e+003 \\ \hline

\end{tabular}

\clearpage

\begin{tabular}{|c|c|c|c|c|c|c|}
\hline No. & $\eta$  &  $\theta$  &  genome size &  $S(\eta,\theta)$
&  gene number  & $N(\eta, \theta)$
\\ \hline

 41 & 0.0910 & 0.2753 & 3284156 & 2.8916e+006 & 3099 & 3.1197e+003 \\ \hline
 42 & & 0.3107 & & & 3524 & \\ \hline
 43 & 0.8100 & 0.2562 & 120000000 & 2.5164e+008 & 18358 & 1.6650e+004 \\ \hline
 44 & 0.1220 & 0.2247 & 4641000 & 4.6505e+006 & 4281 & 4.6032e+003 \\ \hline
 45 & 0.1200 & 0.2852 & 3218031 & 3.2588e+006 & 3145 & 3.1186e+003 \\ \hline
 46 & & 0.2226 & & & 4463 & \\ \hline
 47 & 0.1020 & 0.3149 & 2714500 & 2.4678e+006 & 2067 & 2.4821e+003 \\ \hline
 48 & & 0.2462 & & & 4425 & \\ \hline
 49 & & 0.4102 & & & 1715 & \\ \hline
 50 & 0.1500 & 0.3434 & 4524893 & 2.8080e+006 & 1709 & 2.2966e+003 \\ \hline
 51 & & 0.3185 & & & 2058 & \\ \hline
 52 & & 0.6795 & & & 202 & \\ \hline
 53 & 0.0700 & 0.3495 & 1799146 & 1.6699e+006 & 1874 & 1.8581e+003 \\ \hline
 54 & 0.0920 & 0.3633 & 1643831 & 1.7640e+006 & 1564 & 1.7844e+003 \\ \hline
 55 & 0.9830 & 0.1889 & 3.0000e+009 & 1.0522e+009 & 37229 & 3.7131e+004 \\ \hline
 56 & & 0.3399 & & & 1813 & \\ \hline
 57 & 0.1260 & 0.3358 & 2365589 & 2.5342e+006 & 2266 & 2.2879e+003 \\ \hline
 58 & & 0.2637 & & & 3002 & \\ \hline
 59 & & 0.3320 & & & 2023 & \\ \hline
 60 & & 0.2837 & & & 3652 & \\ \hline
 61 & 0.0970 & 0.2748 & 3011209 & 3.0073e+006 & 2968 & 3.1706e+003 \\ \hline
 62 & 0.0970 & 0.2622 & 2944528 & 3.2293e+006 & 2833 & 3.4342e+003 \\ \hline
 63 & & 0.2999 & & & 4540 & \\ \hline
 64 & & 0.3418 & & & 1687 & \\ \hline
 65 & 0.0800 & 0.3228 & 1751377 & 2.0652e+006 & 1873 & 2.2511e+003 \\ \hline
 66 & & 0.3222 & & & 1869 & \\ \hline
 67 & 0.9500 & 0.1828 & 2.5000e+009 & 8.9214e+008 & 28085 & 3.5960e+004 \\ \hline
 68 & & 0.8092 & & & 80 & \\ \hline
 69 & & 0.2537 & & & 4340 & \\ \hline
 70 & 0.0900 & 0.2451 & 4345492 & 3.4112e+006 & 3906 & 3.7721e+003 \\ \hline
 71 & & 0.5086 & & & 726 & \\ \hline
 72 & 0.1200 & 0.5416 & 580070 & 7.5934e+005 & 484 & 609.2149 \\ \hline
 73 & & 0.5674 & & & 1016 & \\ \hline
 74 & & 0.4804 & & & 686 & \\ \hline
 75 & 0.0860 & 0.4867 & 963879 & 8.4373e+005 & 778 & 802.6126 \\ \hline
 76 & 0.1710 & 0.3407 & 2184406 & 3.2388e+006 & 2065 & 2.4452e+003 \\ \hline
 77 & & 0.3436 & & & 2461 & \\ \hline
 78 & & 0.2513 & & & 3496 & \\ \hline
 79 & & 0.3800 & & & 1909 & \\ \hline
 80 & 0.1060 & 0.2167 & 6264403 & 4.4184e+006 & 5563 & 4.6810e+003 \\ \hline

\end{tabular}

\clearpage

\begin{tabular}{|c|c|c|c|c|c|c|}
\hline No. & $\eta$  &  $\theta$  &  genome size &  $S(\eta,\theta)$
&  gene number  & $N(\eta, \theta)$
\\ \hline

 81 & & 0.2240 & & & 5316 & \\ \hline
 82 & & 0.3236 & & & 1764 & \\ \hline
 83 & & 0.3071 & & & 2065 & \\ \hline
 84 & 0.1320 & 0.3851 & 6397126 & 1.9863e+006 & 2064 & 1.6935e+003 \\ \hline
 85 & 0.1270 & 0.2242 & 5810922 & 4.8078e+006 & 5092 & 4.6687e+003 \\ \hline
 86 & & 0.1953 & & & 7264 & \\ \hline
 87 & 0.1900 & 0.5019 & 1268755 & 1.4540e+006 & 1374 & 912.3452 \\ \hline
 88 & 0.4250 & 0.3018 & 13800000 & 1.8831e+007 & 4987 & 5.4215e+003 \\ \hline
 89 & & 0.4419 & & & 4176 & \\ \hline
 90 & 0.1690 & 0.2845 & 2878084 & 4.4024e+006 & 2631 & 3.4816e+003 \\ \hline
 91 & & 0.3210 & & & 2121 & \\ \hline
 92 & 0.1110 & 0.1809 & 8670000 & 5.5793e+006 & 7894 & 5.9409e+003 \\ \hline
 93 & & 0.3949 & & & 2094 & \\ \hline
 94 & & 0.3350 & & & 1845 & \\ \hline
 95 & & 0.3006 & & & 2977 & \\ \hline
 96 & 0.1300 & 0.3378 & 1564905 & 2.5671e+006 & 1478 & 2.2787e+003 \\ \hline
 97 & 0.0500 & 0.3316 & 1860725 & 1.6375e+006 & 1846 & 1.9943e+003 \\ \hline
 98 & 0.1280 & 0.4457 & 1900521 & 1.3744e+006 & 1031 & 1.1417e+003 \\ \hline
 99 & 0.0700 & 0.5053 & 751719 & 6.8918e+005 & 611 & 688.9768 \\ \hline
 100 & 0.1255 & 0.3336 & 4034065 & 2.5593e+006 & 2736 & 2.3187e+003 \\ \hline
 101 & & 0.2561 & & & 4800 & \\ \hline
 102 & 0.0600 & 0.3362 & 2110355 & 1.6949e+006 & 2044 & 1.9790e+003 \\ \hline
 103 & 0.1440 & 0.2545 & 5175554 & 4.4875e+006 & 4029 & 3.9940e+003 \\ \hline
 104 & 0.1200 & 0.4376 & 2679305 & 1.3711e+006 & 2763 & 1.1815e+003 \\ \hline
 105 & & 0.3221 & & & 6356 & \\ \hline
 106 & 0.1420 & 0.3265 & 4653728 & 2.9448e+006 & 4087 & 2.5139e+003 \\ \hline

\end{tabular}

\end{document}